\documentclass[aps,pra,twocolumn,groupedaddress,superscriptaddress]{revtex4-1}

\usepackage{stmaryrd}
\usepackage{amssymb,amsmath,amsthm,amsfonts,amsbsy}
\usepackage{bm,bibunits,color,chngcntr,epsfig,epstopdf,graphicx,dsfont}
\usepackage{hyperref,lipsum,,makecell,mathrsfs,rotating}
\usepackage[english]{babel}
\usepackage[normalem]{ulem}

\newcommand{\be}{\begin{equation}}
\newcommand{\ee}{\end{equation}}
\newcommand{\bea}{\begin{eqnarray}}
\newcommand{\eea}{\end{eqnarray}}
\newcommand{\ket}{\rangle}
\newcommand{\bra}{\langle}

\newcommand{\I}{\mathds{1}}
\newcommand{\ra}{\rightarrow}

\def\C#1{\mathcal #1}

\newcommand{\T}[2]{\textsf{#1#2}}

\begin{document}
\newtheorem{theorem}{Theorem}
\newtheorem{prop}[theorem]{Proposition}
\newtheorem{corollary}[theorem]{Corollary}
\newtheorem{open problem}[theorem]{Open Problem}
\newtheorem{definition}{Definition}
\newtheorem{remark}{Remark}
\newtheorem{example}{Example}

\title{Theory of quasi-exact fault-tolerant quantum computing and valence-bond-solid codes}

\author{Dong-Sheng Wang}
\affiliation{Institute of Theoretical Physics, Chinese Academy of Sciences, Beijing, China}
\affiliation{Institute for Quantum Computing, University of Waterloo, Waterloo, Ontario N2L 3G1, Canada}
\author{Yun-Jiang Wang}
\affiliation{State Key Laboratory of Integrated Services Networks, Xidian University, Xian, Shaanxi, China}
\author{Ningping Cao}
\affiliation{Department of Mathematics and Statistics, 
University of Guelph, Guelph, Ontario N1G 2W1, Canada}
\affiliation{Institute for Quantum Computing, University of Waterloo, Waterloo, Ontario N2L 3G1, Canada}
\author{Bei Zeng}
\affiliation{Department of Physics, The Hong Kong University of Science and Technology, Clear Water Bay, Kowloon, Hong Kong, China}
\author{Raymond Laflamme}
\affiliation{Institute for Quantum Computing, University of Waterloo, Waterloo, Ontario N2L 3G1, Canada}
\affiliation{Department of Physics and Astronomy, University of Waterloo, Waterloo, Ontario N2L 3G1, Canada}

\date{\today}

\begin{abstract}
In this work,
we develop the theory of quasi-exact fault-tolerant quantum (QEQ) computation,
which uses qubits encoded into quasi-exact quantum error-correction codes (``quasi codes'').
By definition, a quasi code is a parametric approximate code that can become exact
by tuning its parameters.
The model of QEQ computation lies in between the two well-known ones:
the usual noisy quantum computation without error correction
and the usual fault-tolerant quantum computation,
but closer to the later.
Many notions of exact quantum codes need to be adjusted for the quasi setting.
Here we develop quasi error-correction theory using quantum instrument,
the notions of quasi universality, quasi code distances, and quasi thresholds, etc.
We find a wide class of quasi codes which are called valence-bond-solid codes,
and we use them as concrete examples to demonstrate QEQ computation.
\end{abstract}

\maketitle


%
%
%
\section{Introduction}
\label{sec:intro}

A universal quantum computer can realize arbitrary unitary evolution
on a finite-dimensional quantum system and is expected to be more powerful
than classical ones.
The universality requires that any unitary operator in the unitary group
can be realized efficiently to arbitrary accuracy.
The requirement on accuracy is actually a strong one since
qubits carried by physical systems are inevitably affected by noises.
This is usually achieved with quantum error-correction (QEC) codes
leading to the fault-tolerant quantum (FTQ) computation~\cite{Shor96,AB97,KL97,KLZ98,Pre98,NC00}.

Finding good QEC codes is important and many progresses have been made recently.
For QEC, the detected errors are required to be fully or exactly corrected.
The generalization of QEC by allowing approximation has been considered long time ago.
It was observed that~\cite{LNC+97} approximation can be considered as a quantum resource
due to non-orthogonality of quantum states.
Some approximate QEC (AQEC) codes were studied in many settings decades ago~\cite{LNC+97,Pre00,CGS05,RW05},
and recently AQEC is attracting wide attentions partly due to the
connection with holography and quantum gravity~\cite{HNP+17,KK17,FHK+17,BCS+18,BCN+19,GKS+19,FNA+19,WA19}.
It is thus a question whether in general AQEC codes can be used for FTQ computation.
It appears that this is not the case 
since errors are not guaranteed to be corrected completely hence may accumulate quickly.

\begin{figure}[t!]
 \centering
 \includegraphics[width=.7\columnwidth]{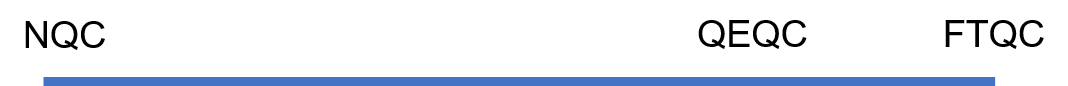}
 \caption{Schematics showing the relation among the three frameworks: 
 noisy quantum (NQ), quasi-exact quantum (QEQ), and fault-tolerant quantum (FTQ) computation. 
 The QEQC introduced in this work is closer to the FTQC.}
 \label{fig:land}
\end{figure}

In this work, we study the framework of quasi-exact quantum (QEQ) computation
introduced recently~\cite{WZO+20},
which allows accurate computation with some kind of AQEC codes.
The two primary notions are quasi-exact QEC codes and quasi-exact universality.
We will call the former as quasi codes, and later as quasi universality, for simplicity.
In essence, a quasi code is a strengthened AQEC code in that
it is parameterized by some scaling parameters the tuning of which can drive it to several exact codes.
That is, in the scaling-parameter space a quasi code has a sort of `phase diagram'
with several exact codes as its fixed points and the rest as approximate regions.
It is clear that a quasi code shall work near fixed points for QEQ computation.
Meanwhile, the quasi universality is a slight weakening of the usual universality
by making a tunable and controllable `cut-off' on the accuracy of gates.
The cutoff effectively induces a coarse-graining structure on the unitary group
by identifying the neighborhood of a gate as the gate itself.
The total number of distinct gates is finite but tunable to approach the usual universality.
In particular, infinitesimal logical gates (measured by some accuracy) are not required in QEQ computation.
The length of a QEQ computation cannot be arbitrarily long,
while its accuracy can be well assessed due to error correction.
Compared with the usual noisy quantum (NQ) computation and FTQ computation,
the model of QEQ computation lies in between while closer to the later,
see Fig.~\ref{fig:land}.

Furthermore, we study more aspects of QEQ computation in details,
including quasi code distance,
the classification of quasi codes,
quasi thresholds,
transversality, etc.
According to the convergence efficiency to exact codes with respect to scaling parameters,
we define two types of quasi distances, strong and weak ones,
and four types of quasi codes.
Note that more types are possible depending on finer criteria.
The threshold theorem has to be modified for quasi codes
since the desired logical error rate cannot be made arbitrarily small.
Instead, we propose to use a pair of thresholds, the physical and logical ones
which both depend on scaling parameters, to specify the fault-tolerance of QEQ computation.

We see that one of the central feature of quasi codes and QEQ computation
is the tunability.
This requires the existence of tunable or controllable parameters,
which actually fitS into usual physical settings that some parameters are sort of controllable,
such as temperature, particle number, chemical potential, etc.
The tunability is a resource so that a user can design the performance of
a QEQ computation in terms of the accuracy, threshold, and code distance.
The idea of using tunable parameters is common in many tasks such as machine learning,
quantum control, etc.
This intuition also applies to the other two models but for QEQ computation
the tunability is achieved by changing some classical scaling parameters.

It is important to find quasi codes to demonstrate QEQ computation.
A natural setting, but not limited to, is the topological phases of matter~\cite{Wen04,ZCZ+15}.
A notable class is based on valence-bond solids (VBS)~\cite{AKLT87},
which is a wide class of models in quantum magnetism and many-body physics.
There are various ways to encode qubits either using bulk ground states, excitations, defects, or edge modes,
and the logical gates they support are not the same. 
Here we define three types: the bulk, edge, and holographic codes,
and compare them through concrete examples.
The VBS codes are quasi-exact due to non-vanishing (usually exponentially decaying) correlation functions,
and weak due to the symmetry-protected topological (SPT) order~\cite{CGW11,SPC11,CGL+12,CGL+13,DQ13a,DQ13b,WAR18,Wang20a}.
Also they are non-stabilizer codes but very similar with stabilizer codes
due to being local and frustration-free.
The edge code behaves similarly with holographic code,
while the bulk code can support some topological logical gates from the SPT order~\cite{WAR18,Wang20a}.
Although the exact distance might be a small constant due to 
the short-range entanglement of ground states~\cite{ZZ14,GKS+19},
the quasi distance could be larger.
With a proper local error model,
we also estimate the quasi-thresholds, which depend on the weak code distances
and they are essentially the same for the three types of VBS codes.

Due to the global symmetry,
the holographic SU($d$) VBS codes are transversal and quasi universal for SU($d$),
which are a class of covariant codes.
It is recently shown that~\cite{FNA+19,WA19,KD20,ZLJ20,YMR+20} covariant codes 
cannot achieve transversal and universal computation efficiently
since arbitrary accuracy will lead to an exponential growth of system size.
This inefficiency originates from the uncertainty relation, 
which manifests as the fundamental limitation on the accuracy of 
estimating an arbitrary unitary operator in quantum metrology~\cite{KD20,ZLJ20,YMR+20}.
Indeed, the usual approach for estimating a gate $U$ employs many copies of it,
which can be viewed as a transversal computation.
For QEQ computation, the cutoff on accuracy will avoid the blow-up
and turn covariant codes into a moderate class of quasi codes~\cite{WZO+20}.
We demonstrate that VBS codes provide a class of covariant codes with novel locality structures. 

This work contains the following parts.
In section~\ref{sec:QEC}, we review the standard theory of
QEC based on channels,
and then we generalize it using instruments,
and we define local QEC instruments in particular for codes with locality structures.
We introduce quasi codes in section~\ref{sec:aqec}.
After a review of AQEC,
we introduce quasi QEC as a strengthened version of AQEC
and discuss some examples.
We study various aspects of QEQ computation in section~\ref{sec:qeqc}.
In section~\ref{sec:VBSC} we study VBS codes and compare three types of them.
We conclude in section~\ref{sec:disc} with comments on some issues and open problems.
For the Appendix,
in section~\ref{sec:cp}, we review completely positive maps,
channels and instruments, and distance measures.
We also introduce two new types of channels:
the subspace CPTP maps and CP but approximate-TP maps,
which can both be employed for QEC.
In section~\ref{sec:aencoding}
we explain how to generalize AQEC by approximate encoding operations,
and study quasi logical gates.
Some future directions are addressed in the conclusion section~\ref{sec:disc}.

\section{Quantum error correction}
\label{sec:QEC}

\subsection{Preliminary}
\label{subsec:QEC}

In this work, we consider finite-dimensional quantum systems.
For a Hilbert space $\C H$, denote the space of bounded,
positive semi-definite, and density operators as $\C B(\C H)$,
$\C P(\C H)$, and $\C D(\C H)$, respectively.
Here we review the standard setting of QEC~\cite{NC00}.
Some basics of quantum channels and instruments are deferred to the Appendix~\ref{sec:cp},
where we also introduce some novel variations.

First, we review the setting of error detection (QED).
A channel $\C N$ on a code subspace $\C C\subset \C D(\C H)$ is detectable when
\be \C P \C N (\sigma) =p \; \sigma,\; p\in [0,1],\; \forall \sigma \in \C C. \ee
If $p=0$, this means $\C N$ causes a complete leakage out of $\C C$.
If $p=1$, this means $\C N$ has trivial action on $\C C$.
With Kraus operators $\{E_i\}$ for $\C N$, the QED condition is
\be P E_i P= e_i P, \ee
and $\sum_i |e_i|^2=p$.

Now we consider the correction (recovery) of a channel, $\C N$, on a code subspace $\C C$.
The correction is by a recovery channel $\C R$ so that
\be \C R \C N (\sigma) =\sigma, \forall \sigma\in \C C. \ee
The condition in terms of Kraus operators $\{E_i\}$ for $\C N$ is
\be P E_i^\dagger E_j P= a_{ij} P, \label{eq:qec}\ee
such that the code matrix $[a_{ij}]:=\rho_*$ is a density operator.

The recovery channel $\C R$ is found by diagonalizing the code matrix $\rho_*$ as $\rho_*=U \rho_D U^\dagger$
for $\rho_D=[d_{k\ell}\delta_{k\ell}]$.
Let $d_k=d_{kk}$, we have $\sum_k d_k=1$.
The basis transformation $U$ can be viewed as a unitary freedom in the Kraus representation of $\C N$
so that the new representation is $\{F_k\}$ and
\be P F_k^\dagger F_\ell P= d_{k\ell}\delta_{k\ell} P.\label{eq:qeco}\ee

The recovery channel $\C R$ is defined by the set of Kraus operators 
\be R_k=\frac{1}{\sqrt{d_k}} P F_k^\dagger \ee 
for $d_k\neq0$, and an additional $R'=\sqrt{\I -P_R}$ for $P_R:=\sum_k R_k^\dagger R_k$.
The operator $P_R$ is a projector on the rotated subspace $\C N(\sigma)$, $\forall \sigma\in \C C$.
Therefore, the operator $R'$ has trivial action on the input $\forall \sigma\in \C C$.

When the set $\{F_k\}$ contains an operator, denoted by $F_0$, which is proportional to identity,
then the condition~(\ref{eq:qeco}) also implies the error detection condition $PF_kP=0$.
This means that $F_k$ causes a complete leakage error out of the code space $P$.
The correction condition~(\ref{eq:qeco}) means
the error operators $F_k$ can be distinguished.
Namely, we can use polar decomposition
$F_k P= \sqrt{d_k} U_k P$, and define $P_k=U_k P U_k^\dagger$ as projectors on the so-called syndrome subspaces
with $P_k P_\ell=\delta_{k\ell}P_k$, and
$PU_k U_\ell P=\delta_{k\ell} P$.
The syndrome subspaces $\{P_k\}$ are orthogonal to the code space $P$ itself.

\begin{figure}
  \centering
  \includegraphics[width=.8\columnwidth]{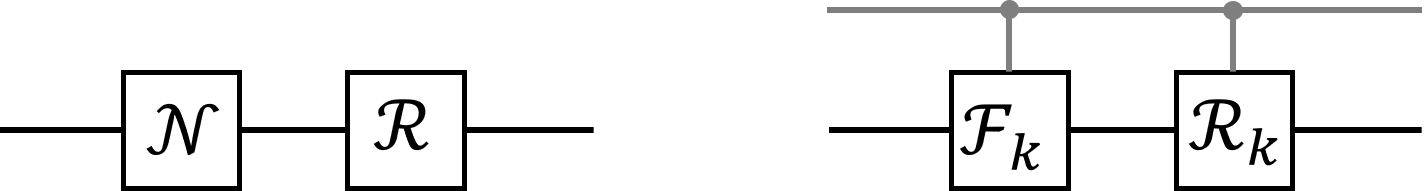}
  \caption{Two ways to realize quantum error correction:
  (Left) the `anonymous' scheme $\C R \C N$ without the identification of syndromes,
  and (Right) the `selective' scheme $\{\C R_k \C F_k\}$ as an instrument
  with a classical register (upper wire).}\label{fig:qecway}
\end{figure}

Due to the orthogonality of syndromes (subspaces),
the error-correction procedure can be done in two ways, see Fig.~\ref{fig:qecway}.
For the first scheme, the procedure is done without the identification of syndromes
by applying $\C R$ after $\C N$ directly.
This leads to
\be
\C R \C N (\sigma)=\sum_{k\ell}R_k F_\ell \sigma F_\ell^\dagger R_k^\dagger=\sigma. \ee
For the second scheme, the procedure is done with the identification of syndromes
and each error $F_\ell$ is corrected by the corresponding $R_\ell$.
This leads to
\be \sum_{\ell}R_\ell F_\ell \sigma F_\ell^\dagger R_\ell^\dagger=\sigma.\ee
The two schemes are equivalent due to the orthogonality of syndromes.

The second scheme can be formulated using the notion of quantum instrument,
which is a set of CP maps (see appendix).
For QEC, we can define error or syndrome instrument $\C N=\{\C F_k\}$,
for $\C F_k$ as the superoperator form of $F_k$,
so that $\C N$ is a channel.
We can also define a recovery instrument
$\C R=\{\C R_k, \C R'\}$ so that $\C R$ is a channel.
Finally, we can define the combined procedure as an instrument
$\C Q=\{\C Q_k\}$ for $\C Q_k= \C R_k \C F_k$ so that $\C Q$ is a sCPTP channel on the subspace $\C C$,
and $\C Q(\sigma)=\sigma$ (see appendix~\ref{subsec:scptp}).
Note the $\C R'$ is unnecessary to define $\C Q$.

There is an important property of the exact QEC above, usually known as `linear Kraus-span (LKS) property':
if a channel $\C N=\{E_i\}$ is correctable by $\C R$,
then any other channel with each of its Kraus operators in the span of $\{E_i\}$ is also correctable by $\C R$.
This fact can be easily verified.
This LKS property simplifies QEC task especially for local QEC:
instead of arbitrary channels, one only needs to consider a channel whose Kraus operators span a local space.
This applies to the case of topological stabilizer codes~\cite{Got98,Kit03}.

\subsection{QEC and QED as instruments}

In the above, the recovery channel is found provided the diagonalization of $\rho_*$.
This can be easily done if $\rho_*$ is a small matrix;
however, this will be difficult if the dimension of it is large
except cases when $\rho_*$ are of particular types.
Also each recovery $R_k$ might be difficult to realize compared with the set $\{E_i\}$.

Here, in terms of quantum instrument, we show that the problem of diagonalization
can be avoided if there are additional algebraic structures in a code.
The notion of QEC instrument applies to the VBS codes studied later in this paper.
For quantum codes supported by large physical systems,
e.g., topological codes,
there are additional structures such as locality.
For instance, for topological stabilizer codes~\cite{Kit03}
there are usually local commuting stabilizers that can be measured
instead of $\{P_k\}$ to identify syndromes.
The locality of stabilizers is usually constant compared with the system size,
which is the size of the code block.
The errors that are being corrected are also local.

Below we present QEC (and QED) in terms of quantum instrument.
This can be either viewed as a restriction of standard QEC when the syndrome is recorded explicitly,
or viewed as a generalization of standard QEC from a single channel to a collection of CP maps,
i.e., from correction of (Kraus) operators to correction of superoperators (CP maps).

A noise instrument $\C N$ is defined as a set of noise CP maps $\{\C N_x\}$ so that
the sum of them is a channel, $\C N$.
It is clear to see that $\C N$ is detectable iff each of them $\C N_x$ is detectable.
This generalizes the standard QED in the sense that the usual operators are generalized to superoperators.

Similarly, we can define recovery instrument and the whole QEC process as an instrument.
\begin{definition}
A QEC instrument $\C Q$ is defined as a set of CP maps $\{\C Q_x\}$
with $\C Q_x=\C R_x \C N_x$ for noise CP maps $\{\C N_x\}$ forming a noise instrument $\C N$
and recovery CP maps $\{\C R_x\}$ so that $\C Q$ is sCPTP on a code subspace $\C C$
and $\C Q(\sigma)=\sigma$, $\forall \sigma\in \C C$.
\end{definition}
Note that we do not require the set $\{\C R_x\}$ as an instrument, i.e. a CPTP map.
Instead, we require the whole process $\C Q$ as a sCPTP map on a code subspace $\C C$.

With the definition above, it is easy to obtain the following:
\begin{prop}
A code space $P$ is protected by a QEC instrument $\C Q=\{\C Q_x\}$ for $\C Q_x=\C R_x \C N_x$
iff $\C Q_x(\sigma)=q_x \sigma$ and $\sum_x q_x=1$, $q_x\in (0,1)$,
for $\C N_x$, $\C R_x$ as CP maps.
\end{prop}
\begin{proof}
It is easy to see that if $\C Q_x(\sigma)=q_x \sigma$, then $\C Q(\sigma)=\sigma$, $\forall \sigma\in \C C$.
If $\C Q(\sigma)=\sum_x \C Q_x(\sigma)=\sigma$,
then as each $\C Q_x(\sigma)$ is positive, it must hold $\C Q_x(\sigma)=q_x \sigma$,
for $\sum_x q_x=1$ with $q_x\in (0,1)$.
\end{proof}
We see that the condition $\C Q_x(\sigma)=q_x \sigma$ is the QEC condition
for each CP map $\C N_x$.
Given $\C N_x$, a recovery CP map $\C R_x$ can be found according to the standard QEC.
Here the index $x$ plays the role of syndrome for superoperators, namely,
it signals which noise CP map $\C N_x$ is identified,
and then which recovery $\C R_x$ shall be applied.
The index $x$ is a classical degree of freedom that induces distinguishability among
the syndromes for the noises $\C N_x$.
However, such a syndrome is not required for the correction of each $\C N_x$ by $\C R_x$.

\subsection{Local QEC instruments}
\label{subsec:lqec}


Now we apply the framework above to the setting when a locality can be defined on a code $\C C$.
The index $x$ in a quantum instrument can also refer to many different degree of freedoms,
such as particle number, super-selection sectors, energy, etc.
For QEC, this information shall be easy to access.
For QEC on stabilizer codes, there are (semi-)local commuting stabilizers that can be measured
to identity syndromes, local Pauli errors, and then the recovery maps.
We define local QEC codes as follows.

\begin{definition}
  A local QEC code $\C C \subset \C D(\C H)$ is defined by a set of check operators $S=\{S_i\}_{i\in \C I}$,
  such that $S_iP=P$, $\forall i\in \C I$, for $P$ as the code projector.
  In particular, when $\C H=\otimes_{n=1}^N \C H_n$,
  the $S_i$ are semi-local check operators such that
  the weight of each of them is $o(1)$ compared with the system size $N$.
\end{definition}
We say $P$ is nonlocal, the check operators $S$ are semi-local, and the errors in QEC are local.
The check operators are often simply referred to being `local', and may not commute with each other.
Although the definition of local codes does not require check operators acting on neighboring sites,
in this work we will implicitly assume a notion of geometrical locality.
This definition applies to many-body quantum systems defined by local frustration-free Hamiltonians,
also it is a special formulation of low-density parity-check codes.
In this setting, a local code can be defined as the low-energy sector of a
local Hamiltonian $H=\sum_n H_n$ with local terms $H_n$,
which usually act on a constant number of neighboring sites
and do not commute with each other in general.
In particular, it applies to the VBS codes we study.

A local QEC instrument is defined as a QEC instrument when the index $x$ refers to locality.
The most important and primary case is when $x$ refers to local sites.
Then the noise we consider is on-site.
Each noise map $\C N_x$ acts on a local site.
However, each recovery map $\C R_x$ may be semi-local since check operators are semi-local,
although they are labelled by $x$.
We shall also keep in mind that in principle $x$ may also refer to semi-local sites or other types of locality.

With respect to local QEC instruments,
the code distance is defined as $d_c=2t+1$ if $t$ independent errors can be corrected.
Let $\rho^{(n)}_{\vec{n}}$ denote $n$-site reduced density operators for any $n$ sites denoted by ${\vec{n}}$.
A code of distance $d_c=2t+1$ shall have $\rho^{(n)}_{\vec{n}}$ for $n\in[1,t]$ all
the same independent of the codewords.
That is to say, the logical information cannot be revealed based on $\rho^{(n)}_{\vec{n}}$.
Note $\rho^{(n)}_{\vec{n}}$ does not have to be a completely mixed state.

With locality, the correction of
high-weight errors is reduced to the correction of low-weight errors.
Noises of the form $\otimes_{n=1}^t \C N_n$ with $n$ as site label,
or mixtures of them $\sum_j p_j \otimes_{n=1}^t \C N_{n,j}$,
can be corrected by the correction of each local noise $\C N_{n,j}$ if distance is $d$.
However, channels of more general forms are not guaranteed to be correctable by local QEC instruments.

Given check operators,
syndrome plays more practical roles in QEC than the actual underlying physical errors:
the syndrome identifies the most likely errors that are detected,
which may differ from the actual physical errors.
The actual errors may be nonlocal;
however, with local syndromes and (semi-)local recovery,
the code can be recovered (semi-)locally.
This can be understood from the LKS property of QEC mentioned above.
In all, we see that our reformulation of QEC in terms of quantum instrument fits in the setting of local QEC very well.

\section{Quasi codes}
\label{sec:aqec}

\subsection{Approximate QEC}
\label{subsec:aqec}

Here we review the standard setting of approximate QEC (AQEC).
A channel $\C N$ on a code $P$ with an encoding isometry $\C V$ is $\epsilon$-detectable if $d(\C V^\dagger \C P \C N \C V, \I)\leq \epsilon$,
for $d$ as a certain distance measure mentioned in section~\ref{subsec:dist}.
A channel $\C N$ on a code $P$ is $\epsilon$-correctable if
there exists a recovery channel $\C R$ such that $d(\C V^\dagger \C R \C N \C V,\I)\leq \epsilon$.
The recovery channel $\C R$ which achieves the smallest $\epsilon$ is not easy to find,
and a convex optimization is needed in general~\cite{RW05,BO10}.

As an extension of (\ref{eq:qec}), there exists $a_{ij}$ and $B_{ij}\in \C B(\C H)$ so that
\be  P E^\dagger_i E_j  P = a_{ij}  P +  P B_{ij} P. \label{eq:aqec}\ee
Furthermore, $P B_{ij} P$ can be chosen to represent the traceless part of $P E^\dagger_i E_j  P$,
so that $a_{ii}\geq 0$ and $\sum_i a_{ii}=1$, $a_{ij}=a_{ji}^*$,
i.e., the matrix $[a_{ij}]:=\rho_*$ is a density operator.
According to~\cite{BO10}, a channel $\C N$ on a code $P$ is $\epsilon$-correctable iff
$ d(\C D+\C B, \C D) \leq \epsilon$,
with $\C D (\rho) := \rho_*^t$,
$\C B (\rho) :=\sum_{ij} \T tr(\rho B_{ij}) |j\ket\bra i|$,
$\rho\in \C C$,
for $d$ as a proper distance measure, 
$\{|i\ket\}$ as states of an `environment'.

The above can be understood using the notion of complementary channel.
For a channel $\C N(\rho)=\sum_i K_i \rho K_i^\dagger$, its complementary channel is
\be \widehat{\C N}(\rho)=\sum_{ij}\T tr(\rho E^\dagger_i E_j)|j\ket\bra i|. \ee
If $\rho$ contains the encoded information,
then $\widehat{\C N}(\rho)$ represents the state of an environment which may contain
some information from $\rho$.
For states in a code space $P$, $\rho=P\rho P$.
For a correctable channel it holds $\widehat{\C N}(\rho)=\rho_*^t=[a_{ij}]^t$,
which is the transpose of the code matrix.
That is to say, $\rho_*$ is the state of the environment
and it is independent of the encoded input state $\rho$.
A correctable channel $\widehat{\C N}$ is a replacement channel.

For AQEC, it becomes $\widehat{\C N}(\rho)=\C D(\rho)+ \C B(\rho)$.
Denote $\widehat{\C N}_0(\rho)=\C D(\rho)=\rho_*^t$.
Then the condition says that if the distance between $\widehat{\C N}$ and $\widehat{\C N}_0$
is upper bounded by $\epsilon$,
then the distance between channels $\C N$ and $\C N_0$ is also upper bounded,
i.e., there exists a recovery channel $\C R$ so that
the distance between $\C V^\dagger \C R \C N \C V$ and $\I$ is upper bounded by $\epsilon$.

By treating quantum instrument as a set of CP maps,
the AQEC of channels also applies to instrument.
Namely, an AQEC instrument $\C Q$ is defined so that $d(\C V^\dagger \C Q \C V, \I)\leq \epsilon$
for $\C Q=\sum_k \C Q_k =\sum_k \C R_k \C N_k$.
However, there is no definite distance bound for each term $\C Q_k$.
The formalism of AQEC can be further generalized when the encoding itself
is also approximate.
This is deferred to Appendix~\ref{sec:aqec_ae}.

Compared with exact QEC, there are several nontrivial facts about AQEC.
First, the recovery channel is no easier to find in the complementary picture
than in the original picture, in general.
Second, as the task of AQEC is channel-oriented,
the recovery scheme is not universal, i.e.,
it is not guaranteed to work for other channels acting on the code.
The error bound for a noise channel may be different from the error bound for another channel
given a fixed recovery scheme.


\subsection{Quasi-exact QEC}
\label{subsec:qqec}

Now we introduce a special type of AQEC codes, the quasi-exact AQEC codes,
or simply, quasi codes.
Roughly speaking, an approximate code is a quasi code if it can get arbitrarily close to an exact code
in certain ways.
In other words, a quasi code can be viewed as a perturbation of an exact code.
The transition between an exact code and an approximate code is done by tuning of a certain parameters
that are inherent to a quasi code.
The motivation for introducing quasi codes is multiple.
The QEC errors of approximate codes have to be small enough in order to do
accurate quantum computing.
Also in practice there are usually some parameters that can be controlled and tuned
so that the performance of a code can be improved.
We define quasi codes as follows.

\begin{definition}[Quasi codes (broad)]
A quasi code $\C C(\vec{\lambda})$ is a family of
$\epsilon(\vec{\lambda})$-correctable approximate codes with a recovery scheme $\C R(\vec{\lambda})$,
such that each code is defined with fixed values of $\vec{\lambda}$,
which is a vector of a finite number of real scaling parameters,
and $\epsilon(\vec{\lambda})\ra 0$
at some points in the parameter space of $\vec{\lambda}$.
We assume that $\epsilon$ is a smooth function of $\vec{\lambda}$.
\end{definition}

A quasi code can be labeled as $\C C(\vec{\lambda})$,
or by a tuple
\be \C C(\vec{\lambda}):= \bra \C V, \C N, \C R, \vec{\lambda}, \epsilon \ket,\ee
while $\vec{\lambda}$ may not exist for general approximate codes.
The parameters $\vec{\lambda}$ are assumed to be real and named as \emph{scaling parameters},
since this may been seen as an analog (but not the same) of critical theory of phase transition
in the sense that a critical point can be reached by tuning of some scaling parameters,
such as interaction strength, temperature, etc.
Indeed, in the parameter space of $\vec{\lambda}$ we can treat exact codes as critical points,
and other non-critical regions as for approximate codes,
while a quasi code serves as a general reference to the whole `phase diagram.'
However, we will use `quasi code' to refer to the vicinity of exact codes,
since these are the regions that are useful for accurate quantum computing.
We assume there are a finite number of scaling parameters.
We say a quasi-to-exact (QTE) limit is a limit by tuning some parameters
of $\vec{\lambda}$ so that $\epsilon(\vec{\lambda}) \ra 0$ is achieved.
The number of limits is arbitrary.
There are diverse choices of $\vec{\lambda}$ defining different quasi codes.
These parameters could be the system size, local dimension, temperature, density of states,
chemical potential, energy, particle number, etc
that are shall be easy to access and controllable in practice.
Also numerical optimization might be involved to find suitable values of scaling parameters
for general quasi codes.

The encoding operation $\C V$ may be an exact or approximate isometry (see section~\ref{sec:aqec_ae}).
We say an $\epsilon_e$-approximate isometry is a quasi isometry if $\epsilon_e$
is a function of $\vec{\lambda}$ and $\epsilon_e \ra 0$ in some limits of $\vec{\lambda}$.
The encoding inaccuracy $\epsilon_e$ will add to the total correction error.
Here we assume the error $\epsilon(\vec{\lambda})$ already account for it if the encoding is approximate.

The noise model $\C N$ and recovery scheme $\C R$ also depend on $\vec{\lambda}$.
The dependence of $\C N$ on $\vec{\lambda}$ shall be `minimal':
it shall be almost irrelevant to $\vec{\lambda}$ so that a quasi code distance $d_c(\vec{\lambda})$ can be well-defined.
For instance, $\C N$ can be 1-local errors and its locality shall be fixed,
irrelevant to $\vec{\lambda}$.
The noise model shall remain to be nontrivial when the quasi code approaches to be exact.
On the contrary, the recovery scheme $\C R$ may depend on $\vec{\lambda}$ significantly.
For each value of $\vec{\lambda}$,
an optimization is needed to find an optimal recovery $\C R(\vec{\lambda})$.
The operation $\C R(\vec{\lambda}) \C N(\vec{\lambda})$ forms the QEC procedure for channels,
while can also be extended to instruments.

Finding quasi codes could be difficult due to the optimization tasks that are involved.
However, this could be made easier if we modify the definition of quasi codes
by using a fixed recovery scheme.
After all, we expect to use codes that are perturbations of exact codes,
for which the recovery scheme is well established (see section~\ref{sec:QEC}).


\begin{definition}[Quasi codes (restricted)~\cite{WZO+20}]
A quasi code $\C C(\vec{\lambda})$ is a family of
$\epsilon(\vec{\lambda})$-correctable approximate codes with the recovery scheme for exact codes,
such that each code is defined with fixed values of $\vec{\lambda}$,
which is a vector of a finite number of real scaling parameters,
and $\epsilon(\vec{\lambda})\ra 0$
at some points in the parameter space of $\vec{\lambda}$.
We assume that $\epsilon$ is a smooth function of $\vec{\lambda}$.
\end{definition}

\begin{figure}[t!]
  \centering
  \includegraphics[width=.8\columnwidth]{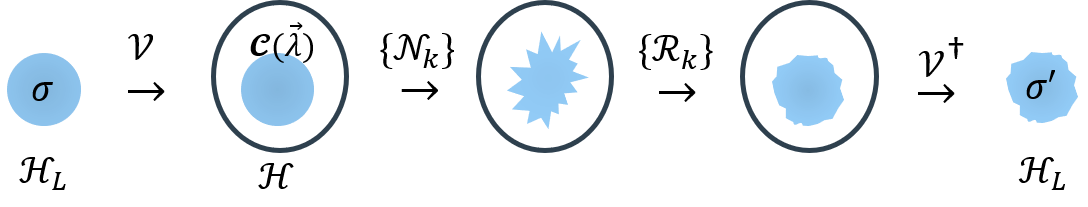}
  \caption{Schematics of quasi codes with encoding $\C V$, noise model $\{\C N_k\}$,
  recovery scheme $\{\C R_k\}$, and the logical states approach the original ones
  $\sigma'\rightarrow \sigma$ by tuning scaling parameters $\vec{\lambda}$.
  For exact codes, the logical states are exactly recovered,
  while for general approximate codes there may not be scaling parameters in order to
  approach exact codes.
    }\label{fig:quasic}
\end{figure}

Our definitions above also apply to QED and extend to the setting of instrument.
This restricted definition is justified by treating a quasi code
as a slight perturbation of an exact code,
which is natural from a physical point of view.
See Fig.~\ref{fig:quasic} for an illustration of quasi codes.
An exact code carried by a physical system may not be actually exact
due to practical inaccuracy and noises.
On the other hand, by extending exact codes to quasi codes
there are new space and opportunity to construct new class of codes.

Furthermore, there exists an explicit formula of the recovery error using trace distance on Choi states (see appendix~\ref{sec:cp}).
With exact encoding, we find
\be D_t(\omega_{\C V^\dagger \C Q \C V}, \omega) = \frac{1}{2d_L} \sum_{kl}d_k \T tr (\hat{B}^\dagger_{kl}\hat{B}_{kl}), \ee
which is a function of $d_k$ and $\beta_{kl}:=\T tr (\hat{B}^\dagger_{kl}\hat{B}_{kl})$,
and also the cardinality of the set $\{k\}$,
 i.e., the size of the `environment', $d_E$.
For local codes, the dimension $d_E$ is a polynomial of
the system size $N$ and local site dimension $d_q$, namely,
$d_E=N^{c_1}d_q^{c_2}$ for some positive constants $c_1$, $c_2$.
In order for it being quasi exact,
$D_t$ has to converge to zero for some values of $\vec{\lambda}$.
That is to say, $d_k$ and $\beta_{kl}$ shall depend on $\vec{\lambda}$
and show universal decay behaviors with some scaling parameters~\cite{WZO+20}.
With quasi encoding, we find
\be D_t(\omega_{\C V^\dagger \C Q \C V}, \omega_{\C V^\dagger \C V})
= \frac{1}{2d_L} \sum_{kl}d_k \T tr (\breve{B}^{\dagger}_{kl}\breve{B}_{kl}), \ee
for $\breve{B}_{kl}:=\breve{\I}\hat{B}_{kl}\breve{\I}$ with $\breve{\I}:=V^\dagger V=\I+\Delta$ due to quasi encoding.
Note we ignore terms with $\Delta^2$ and its higher orders.
The $\breve{\I}$ affects the recovery operations, which depend on the quasi projector $P$,
hence enters the trace distance.
The distance above will add to the encoding error.
In order for a quasi code with quasi encoding becomes an exact code,
the encoding itself at least has to converge to be exact.

Other equivalent distance measures can also be used which may not have explicit forms.
However, what we observe is that the distance is a measure of the average size of these operators $\hat{B}_{kl}$,
instead of each individual ones.
This means that each error-correction round is for the noise channel (or instrument),
which is a collection of Kraus operators (or CP maps),
instead of individual Kraus operators (or CP maps).
For instance, for the correction of local noises on a code supported by a many-body quantum system,
an error-correction round is a statistical average of the correction of local noises at various locations.

The recovery scheme that works for exact codes may not work well for approximate codes in general,
leading to an optimization problem, while it is only slightly disturbed for quasi codes.
From section~\ref{sec:QEC}, we see that the operator $R$ is a projector on the rotated space $\C N\C C$,
and the operator $R'=\sqrt{\I-R}$ never needs to be implemented.
For quasi codes, $R$ is no longer a projector but close to it, i.e., $R$ is a quasi projector.
The QEC instrument $\C V^\dagger \C Q \C V$ acting on $\C H_L$ is not TP but aCPTP (see appendix~\ref{subsec:aCPTP}),
with perturbation terms $\T tr(\hat{\sigma}\Delta)$ and $\T tr (\breve{B}^{\dagger}_{kl}\breve{B}_{kl})$.
When a QTE limit is taken, these terms will vanish,
making the QEC process as an identity channel acting on $\C H_L$,
i.e., a sCPTP channel acting on $\C H$.

The definition of quasi codes in this work is not channel-oriented, i.e.,
a quasi code is not designed for a special error model
(e.g., depolarizing channel or amplitude-damping channel).
Instead, it is more of the spirit of exact codes which possess the LKS property~(from Sec.\ref{subsec:QEC}).
The correction of a noise channel $\C N$ on a quasi code shall also apply to other noises
that are in the span of it, which is the LKS property that holds for exact codes but not for approximate codes in general.
Actually, this is also partly a motivation for the introduction of quasi codes,
as mentioned in section~\ref{subsec:aqec}.
For quasi codes, what is more crucial for the recovery error $\epsilon$ is not its value,
instead it is the \emph{scaling behaviors} with respect to $\vec{\lambda}$.
Given $\{E_i\}$ for the channel $\C N$, hence the set of $\beta_{kl}$ and $d_k$ as functions of $\vec{\lambda}$,
any other channel with each Kraus operator in the span of $\{E_i\}$
will lead to the new set of $\beta_{kl}'$ and $d_k'$,
which are functions of $\vec{\lambda}$.
Therefore, the scaling behaviors of the recovery error shall be preserved against the linear-span operation,
hence the LKS property holds for quasi codes.

\subsection{Examples}
\label{subsec:ex}

The notion of quasi codes apply to some well-known approximate codes in literature. 
Actually, many approximate codes are quasi codes.
We introduce the term `quasi codes' to highlight 
the strong requirement of universality for accurate quantum computing,
and the role of scaling parameters, which are common in physics.
Here we mention briefly two notable classes: 
the gapped quantum phases of matter~\cite{Wen04,ZCZ+15}, 
in particular, anyons~\cite{NSS+08},
and continuous-variable systems such as 
the cat codes (and its generalizations)~\cite{MLA+14,MSB+16,AND+18}.

The ground subspaces of gapped quantum phases of matter are natural to encode qubits,
and generically the codes are quasi-exact.
A gapped phase is often defined by a family of Hamiltonians $H(\vec{\lambda})$ in the thermodynamic limit,
with some universal features common in the phase,
for parameters $\vec{\lambda}$ that could be interaction strength, external field, etc. 
The exact forms of ground states $|G(\vec{\lambda})\ket$ are in general not analytic,
while there are often exactly solvable points for certain $\vec{\lambda}_0$.
The codewords $|G(\vec{\lambda}_0)\ket$ on their own might form an exact or quasi-exact code depending on the error model,
while using $|G(\vec{\lambda})\ket$ will lead to further approximations of $|G(\vec{\lambda}_0)\ket$.
A finite system size is also a source of approximation.
Quantitative study of the quasi QEC features can be carried out using matrix-product states~\cite{PVW+07,Sch11},
quantum field theory, or the geometric method based on 
joint numerical range (JNR)~\cite{buoni1978joint,xie2020observing,spitkovsky2018signatures,chen2016geometry}.

The braiding of non-Abelian anyons,
as excitations of gapped topological systems, can enable universal quantum computing.
There are in general exponentially decaying correlations among anyons,
which makes the code quasi-exact.
Indeed, anyons have to be far away from each other compared with the correlation length.
For instance, a qubit can be encoded by 1D Majorana wire~\cite{SFN15}
with Majorana zero modes (MZM) at each end.
Two MZM form a fermion,
for which a basis is formed by the occupied $|1\ket$ and unoccupied $|0\ket$ states.
The states have overlap $e^{-N/\xi}$
which vanishes when the separation between the modes $N\ra \infty$,
given a finite correlation length $\xi$.


Codes with continuous variables (or oscillators) often involve parameters,
such as squeezing, photon number, etc that can be controlled externally,
hence serving as natural systems for quasi codes. 
The class of cat codes usually employs coherent states of photons,
or bosons in general, and their superposition for the encoding.
For instance, a logical qubit can be encoded as
$|+\ket=|\alpha\ket$, $|-\ket=|-\alpha\ket$,
and $\bra-|+\ket=e^{-\alpha^2}\ra 0$ if $|\alpha|\ra\infty$.
The errors that are natural for cat codes include loss or gain
by a certain orders of $\hat{a}$ or $\hat{a}^\dagger$,
and dephasing errors of the form $e^{i\theta \hat{n}}$.
The violation of the condition~(\ref{eq:qec})
is typically by a term of order $O(\alpha^{c_1} e^{-c_2 \alpha^{c_3}})$,
with constants $c_1,c_2,c_3$~\cite{AND+18}.



\section{QEQ computation}
\label{sec:qeqc}
\subsection{Universality}
\label{subsec:uni}

Now we study computation with quasi codes.
Here we review the definition of universality briefly~\cite{Fey82,BV97,Kit97,Div00,Sho02,LJL+10},
although there appears to be notions that differ slightly;
e.g., some definitions may rely on ancilla and some not.
We employ the most simplest and common definition here.

\begin{definition}[Universality]
  A computation on $n$ qubits, for any integer $n$, is universal if
  any unitary gate $U\in \text{SU}(2^n)$ can be efficiently simulated to arbitrary accuracy.
\end{definition}
The first thing to notice is that the group SU($2^n$), or SU($d$) for any integer $d$,
is uncountably infinite.
Some unitary matrices may contain numbers that are not efficiently computable,
hence shall be replaced by computable ones~\cite{BV97}.
The group SU($2^n$) shall be replaced by such a modified or punctured version.
For simplicity, we still refer to the group SU($2^n$) for the definition of universality.

The efficiency refers to the scaling of the overhead (cost) as a function of the problem size.
However, there is no unique measure of `problem size'.
A usual choice is to use the amount of bits that represent the input.
The cost is measured by space and time, e.g., circuit size or depth.
The accuracy is measured by operator-norm distance or its equivalences.
If an accuracy parameter $\epsilon$ is given as an input for an algorithm,
then an efficient simulation cost shall scales as poly$\log\frac{1}{\epsilon}$.
The log is there because decimal numbers are represented as binary numbers, i.e.,
strings of bits.
A universal quantum computer can run algorithms that have arbitrary accuracy given as input,
such computations are terms as ``algorithmic'' (in the setting of quantum simulation)~\cite{San13,WS15},
compared with others such as analog simulators or emulators.
An NQ computation or NISQ device is normally non-algorithmic
since the computation accuracy is either not given as input or not assessable.

A gate set $S$ is called universal if any $U\in \text{SU}(2^n)$ 
can be efficiently simulated to arbitrary accuracy
by a finite product of gates from $S$.
There are a few well-known gate set: $\{H,T,CZ\}$, $\{H,CCZ\}$, and $\{H,CS\}$,
for $H$ as Hadamard gate, $T$ as $Z^{1/4}$ gate,
$CZ$ as controlled-$Z$ gate, $CCZ$ as controlled-$CZ$ gate,
and $CS$ as controlled-$S$ gate, for $S=T^2$ as the phase gate.
A finite universal gate set generates a dense subset of SU($2^n$),
with the group itself as the closure of it.

The Solovay-Kitaev theorem and algorithm~\cite{NC00,DN06} prove that
there exists a classical algorithm based on balanced group commutator with runtime $t\in O(\log^{c_t}\frac{1}{\epsilon})$
that designs a quantum circuit with size $s\in O(\log^{c_s}\frac{1}{\epsilon})$,
for some known numbers $c_t,c_s\in(1,4)$.
The lower bound for the circuit cost is $\Omega(\log\frac{1}{\epsilon})$,
which can be achieved with methods that reduces the problem to the compiling of $Z$-rotations~\cite{KMM13,Sel12}.
In the above, the accuracy parameter $\epsilon$ only results from gate compiling,
hence referred to as the \emph{compiling error}.
Correspondingly, the cost is the \emph{compiling cost}.
It assumes that each gate can be exactly performed, and the identity gate $\I$ is exact.

In practice, there are various sources of decoherence leading to noisy memory and gates.
In the setting of FTQ computation,
noises are corrected based on QEC so that
arbitrarily long computation can be done provided the noise strength is below a threshold,
which is essentially the content of the threshold theorem~\cite{NC00}.
We refer to the cost for encoding qubits as the \emph{coding cost},
and accuracy parameter as the \emph{coding error} if the error-correction is not exact.

Without powerful QEC,
arbitrarily long computation can not be done.
Fortunately, there is no need to always pursue arbitrarily long computation in practice.
Primary criteria has been the standard for implementations~\cite{Div00},
and there are also requirements for reliable quantum simulators~\cite{Fey82,Llo96,CZ12},
and in general, the NISQ devices~\cite{Pre18}.
These devices or computers carry out NQ computation and do not require explicit error corrections,
hence cannot achieve arbitrary accuracy or arbitrarily long computation,
but still useful to run some algorithms or demonstrate quantum features.


\subsection{Quasi-exact universality}

Here we analyze the concept of quasi universality introduced in Ref.~\cite{WZO+20} in more details.
The computation with quasi codes shall fit in between NQ computation and FTQ computation,
since the error correction of quasi codes cannot be exactly done. 
Uncorrectable noises will accumulate and induce nontrivial logical errors as the size of computation grows.
Furthermore, we require that the usual (exact) universality can be approached
from quasi universality by tuning some controllable parameters.

We introduce quasi universality in the following way.
First, we employ a `coarse-graining' scheme to convert the group SU($2^n$) into a set.
Namely, a gate $U\in \text{SU}(2^n)$ and its neighbors within a distance $\eta$ to it is treated the same as it logically,
forming a `gate cell',
and the group SU($2^n$) can be partitioned into a finite collection of gate cells.
The cells shall not overlap with each other,
and the partition is not unique but can always be chosen and kept fixed.
This partition is similar with the `net' construction in the proof of Solovay-Kitaev theorem~\cite{NC00},
e.g., the lookup table, 
but here the sizes of cells depend on $\eta$ and may be different from each other.

It could be difficult to construct the gate-cell structure for the group SU($2^n$) as a whole,
so we can reduce the task to the smaller group SU(2) and their couplings.
This can be understood for the case with the gate set $\{CZ,R\}$, and $R$ represents the set of single-qubit gates.
Any gate $U\in \text{SU}(2^n)$ can be decomposed into a product of $CZ$ and qubit gates.
Any qubit gate can be written as $U(\theta,\vec{n})=e^{i\theta \vec{n}\cdot \vec{\sigma}}$, ignoring global phases.
From $U(\theta,\vec{n})=U(\theta-\pi,\vec{n})$,
the set of qubit gates can be represented as a radius-$\frac{\pi}{2}$ ball.
A gate-cell structure can be defined from a tessellation of the ball, e.g., the cubic lattice.
Given the inaccuracy $\eta$, the side length of each small cube is $\eta/\sqrt{3}$.
Such a partition is uniform as all gate-cells have the same size.

\begin{figure}[t!]
  \centering
  \includegraphics[width=.8\columnwidth]{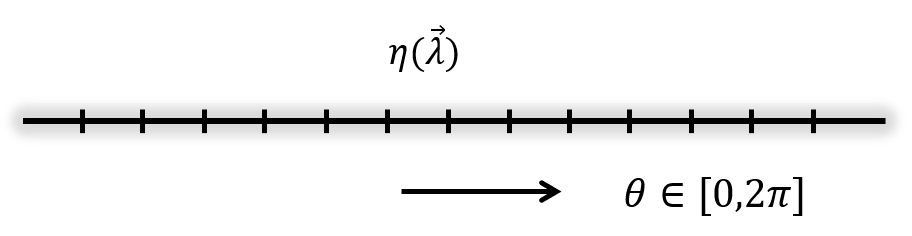}
  \caption{The schematic gate-cell structure for the group $U(1)$ parameterized
  by a single rotation angle $\theta\in [0,2\pi]$ with accuracy $\eta(\vec{\lambda})$.
  The gate compiling cost refers to the realization of all segments,
  and the coding cost refers to the convergence from the digitized set to the continuum by tuning $\vec{\lambda}$.}\label{fig:u1}
\end{figure}

The simplest case probably is for rotations $Z(\theta)=e^{i\theta Z}$, $\theta\in [0,2\pi]$,
if any qubit gate $U\in \text{SU}(2)$ is decomposed as
\be U=Z(\theta_1)HZ(\theta_2)HZ(\theta_3), \ee
and we only need to partition the circle $[0,2\pi]$ into a set of segments, see Fig.~\ref{fig:u1}.
It is natural to assume that all segments have the same size, $\eta(\vec{\lambda})$.
A quasi code which is quasi universal for U(1) rotations is first required to realize all segments efficiently, and then when $\vec{\lambda}$ is tuned,
the segment size decreases and their total number increases,
approaching the group U(1).
The cost for covering all segments is the gate compiling cost,
while the cost to achieve the limit $\text{U(1)}_{\eta(\vec{\lambda})}\rightarrow \text{U(1)}$ is the coding cost.

When using a discrete gate set, e.g., $\{H, T\}$,
any gate $U\in \text{SU}(2)$ needs to be decomposed in terms of them.
Each of $CZ$, $H$, and $T$ may have its own inaccuracy $\eta_{CZ}$, $\eta_{H}$, and $\eta_{T}$.
A qubit gate, as a product of $H$ and $T$, and also $\I$,
will have larger gate cell whose size depends on $\eta_{H}$, $\eta_{T}$, and $\eta_{\I}$.
Such a partition will be non-uniform.
As the result, the whole group SU($2^n$) is reduced to a finite set of gate cells of various sizes,
each as a distinct logical gate.

The gate-cell construction above applies both to the NQ computation and QEQ computation.
The cell size can be viewed as the gate infidelity,
or other equivalent distance measures, which are commonly used in experiments
and can also be improved based on experimental technique.
For QEQ computation,
the gate cell size is also tunable in order for it being exact.
For simplicity, we denote the gate-cell size parameter by a single $\eta$,
which is a function of $\vec{\lambda}$.
Compared with compiling error, here $\eta$ serves as the coding error.
Denote the logical gate set as $\text{SU}(d)_{\eta(\vec{\lambda})}$.

\begin{definition}
[Quasi universality~\cite{WZO+20}]
A computation on a quasi code $\C C(\vec{\lambda})$ with accuracy $\eta(\vec{\lambda})$
is quasi universal for a unitary group SU($d$) if
the coarse-grained set $\text{SU}(d)_{\eta(\vec{\lambda})}$ can be realized efficiently,
and the group SU($d$) can be approached when $\eta(\vec{\lambda})\rightarrow 0$ in a certain QTE limit.
\end{definition}

Here the efficiency refers to the scaling of the cost with respect to parameters of $\text{SU}(d)_{\eta(\vec{\lambda})}$,
such as $d$.
Given a quasi code $\C C(\vec{\lambda})$ and a QEC instrument on it,
the QEQ computation cannot achieve arbitrary accuracy.
It has been estimated that~\cite{WZO+20} the accuracy parameter is lower bounded by
$m\eta$, for $m$ as the size of the computation,
and $\eta$ as the smallest cell size.
As a result, QEQ computation cannot perform algorithmic computation or simulation~\cite{San13,WS15},
for which the accuracy is the input of an algorithm;
however, the accuracy can be estimated and improved based on the quasi code being used.
In other words, a quasi code induces a tunable and controllable `cut-off' on accuracy,
and QEQ computation is weaker than FTQ computation which allows arbitrary accuracy,
but stronger than NQ computation for which the accuracy may not be accessible.
In this sense, the QEQ computation is ``quasi algorithmic.''
We refer to a QEQ computation with the smallest gate-cell size $\eta$ as ``$\eta$-QEQ computation.''
Furthermore, the scaling parameters $\vec{\lambda}$ are given as resources,
and the coding cost relating to the tuning of scaling parameters $\vec{\lambda}$
will be counted separately
from other cost independent of $\vec{\lambda}$,
such as the compiling cost in terms of space and time for classical or quantum circuits.

\subsection{Classification and classical simulation}

Here we provide a brief classification of quasi codes and QEQ computation.
The method is based on local states, or equivalently, the AQEC condition.
Recall that $\rho^{(n)}_{\vec{n}}$ denote an $n$-site reduced density operators.
We find there are two distinct types, which we term as \emph{weak} and \emph{strong} quasi codes:
\begin{enumerate}
\item Strong quasi codes: Each $\rho^{(n)}_{\vec{n}}$ converges
        to the same state independent of the codewords in a QTE limit.
\item Weak quasi codes: Only the statistical average $\overline{\rho}^{(n)}:=\sum_{\vec{n}} p_{\vec{n}} \rho^{(n)}_{\vec{n}}$ converges
to the same state for each $n$ independent of the codewords in a QTE limit.
Here the average is over locations with the number of sites $n$ kept fixed,
and $p_{\vec{n}}$ denotes probability, which depends on the system size, $N$.
\end{enumerate}
The inaccuracy, i.e., recovery error, of a quasi code is measured by a distance
between the QEC channel $\C Q$ and identity $\I$.
As a channel, $\C Q$ depends on the statistical average $\overline{\rho}^{(n)}$,
so the weak quasi codes is a valid family of quasi codes.
Therefore, we can define both weak and strong quasi code distances.

The distinction between the two types of code distance is crucial for quantum codes,
while may be unnecessary for some classical codes.
For instance, the repetition code encodes 0 (1) as a collection of 0 (1),
and there is no sense of worst case or average case for flip errors.
The weak code distance can also be used for exact quantum codes.
For example, the distance of toric code is $\sqrt{N}$ for $N$ as system size.
However, only a small fraction of weight $t=\lceil 1+(\sqrt{N}-1)/2 \rceil$ noises that are close enough geometrically
can cause a logical error among all weight-$t$ noises.
The weak quasi code distance for the toric code is larger than $\sqrt{N}$.
If dilute noises are common than dense ones,
the toric code will perform better than what the strong distance allows.
This agrees with the result that toric code has a high threshold against loss errors
based on percolation theory~\cite{SBD09}.
This reveals that the weak code distance is a valid notion for quantum codes.

For a quasi code with a QEC instrument $\C Q=\sum_x \C Q_x$,
the inaccuracy induced by each $\C Q_x$ may depend on $x$ or the system size, or both.
For local codes, the index $x$ refers to locations, $n$.
Each inaccuracy can take various forms.
For simplicity, here we emphasize four types:
exponentially decay $x^{-n}$ or $x^{-N}$, power-law decay $n^{-\alpha}$ or $N^{-\alpha}$,
for some parameters $x$ and $\alpha$ with $|x|>1$ and $\alpha>0$.
It is obvious to see that the types depending on $n$ ($N$) are weak (strong) quasi codes.

These four types of quasi codes above can be realized in topological phases of matter~\cite{Wen04,ZCZ+15}.
Topological (TOP) orders are long-range entangled,
while symmetry-protected topological (SPT) orders are short-range entangled.
Exponentially decaying correlation functions are common for gapped phases,
while power-law decaying correlation functions are common for gapless phases.
The correspondence is shown in Table~\ref{tab:classes} with some example systems.
As a consequence, there are at least four families of QEQ computation
which are quasi universal for a certain unitary group.

\begin{table}[b!]
    \centering
    \begin{tabular}{|c|c|c|} \hline \Xhline{3\arrayrulewidth}
             & Strong           & Weak  \\ \hline
 Exponential & gapped TOP order      & gapped SPT order \\ \hline
 Power-law   & gapless TOP order     & gapless SPT order \\ \hline \Xhline{3\arrayrulewidth}
    \end{tabular}
    \caption{Four families of quasi codes realized in topological phases of matter.
    Some examples are fractional quantum Hall liquids and quantum spin liquids with gapped TOP order~\cite{Wen04},
    various VBS models with gapped SPT order~\cite{CGW11,CGL+12,CGL+13},
    some quantum dimer model~\cite{RK88} and defect model~\cite{LAS20} with gapless TOP order,
    and both bosonic and fermionic models with gapless SPT order~\cite{KB15,SPV17}.
    }
    \label{tab:classes}
\end{table}

To see the differences of them,
let us consider the convergence from quasi universality to universality.
Due to the gate-cell structure,
the number of distinct logical gates, $N_G$, is not infinite;
instead, it scales with the inverse of the inaccuracy.
It is easy to see that it scales with $N$ for weak quasi codes,
and as $x^{N}$ or $N^{\alpha}$ for strong quasi codes.
The coding cost, when measured by the system size $N$,
scales as $\log\frac{1}{\epsilon}$ for $(x^{-N})$-type strong quasi codes,
while $\frac{1}{\epsilon}$ for weak quasi codes and $(N^{-\alpha})$-type strong quasi codes.
As the order of the permutation group $S_m$ on $m$ bits is $m!$,
it is therefore necessary to have at least $m!$ invertible functions, hence unitary gates on $m$ qubits,
if each bit is substituted by a qubit.
The factorial $m!$ grows comparably or even faster than $2^m$ asymptotically.
As a result, we conclude that the $(x^{-N})$-type strong quasi codes
leads to an efficient convergence, in terms of coding cost,
from quasi universality to universality.
We shall note that the gapped TOP orders is only one notable examples of $(x^{-N})$-type strong quasi codes,
and it is expected that other kinds of phases of matter
or other quantum systems may also be in this family.

The convergence from quasi universality to universality
is irrelevant with classical simulatability of QEQ computation.
Suppose a $\eta$-QEQ computation is quasi universal for SU($2^n$),
can this be efficiently simulated on a classical computer?
A circuit performing QEQ computation may be small, especially for weak quasi codes,
which means the circuit depth may be small.
However, due to the quasi universality,
a unitary circuit $U$ can drive the usual starting state $|00\cdots 0\ket$
across the whole Hilbert space of $n$ qubits,
although it cannot reach every point.
This means that the intermediate states could contain extensive amount of entanglement.
In terms of matrix-product states~\cite{PVW+07},
this is to say that the bond dimension could grow exponentially with $n$,
hence the circuit cannot be efficiently simulated classically.
Therefore, quasi universal quantum computers shall be more powerful than classical computers, in general.

\subsection{Thresholds}

The threshold theorem states that if the physical error rate, $\epsilon_p$, given a proper error model and
measure of errors, is below a threshold value, $\epsilon_p^*$,
the logical error rate, $\epsilon_l$, can be reduced to be arbitrarily small so that
arbitrarily long FTQ computation can be done~\cite{NC00}.
The fault-tolerance is achieved based on QEC.
The threshold value $\epsilon_p^*$ may depend on the codes and decoding algorithms.

With quasi codes, the threshold theorem has to be modified since QEQ computation cannot be arbitrarily long or accurate.
\begin{definition}
[Quasi thresholds]
If the physical error rate, $\epsilon_p$, given a proper error model and
measure of errors, is below a quasithreshold value, $\epsilon^*_p$,
the logical error rate, $\epsilon_l$, can be reduced below $\epsilon^*_l$ so that
$\epsilon^*_l$-QEQ computation can be performed,
and $\epsilon^*_l$ can approach zero in QTE limits.
\end{definition}
A quasithreshold theorem is readily proved by observing that it is a weaker version of the usual (exact)
threshold theorem.
The quasithreshold contains a pair of parameters $(\epsilon^*_p,\epsilon^*_l)$ instead of just one.
The quasithreshold values $\epsilon^*_p$ and $\epsilon^*_l$ may depend on the quasi codes,
in particular, the scaling parameters $\vec{\lambda}$, and decoding algorithms.

It might be the case for some quasi codes that $\epsilon^*_p$ approaches to zero in QTE limits.
To allow a non-zero $\epsilon^*_p$,
the quasi code distance $d_c$ of a quasi code shall get larger when the dimension of $\C H$ increases,
e.g., under concatenation,
for $\C H$ as the Hilbert space of the physical system.
Therefore, the quasi code distance $d_c$ of a quasi code needs to be defined properly.

\begin{figure}
  \centering
  \includegraphics[width=.8\columnwidth]{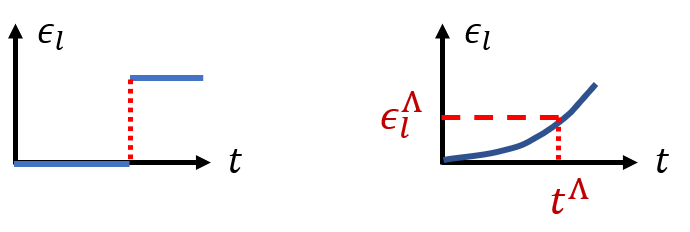}
  \caption{Schematics to show the difference between exact code distance (Left)
  and quasi code distance (Right).
  A tunable cutoff $\epsilon_l^\wedge$ shall be implemented to induce
  a proper quasi code distance $d_c=2t^\wedge+1$.}\label{fig:dist}
\end{figure}

For block codes, the exact code distance $d_c=2t+1$ is defined
when a code can exactly correct up to $t$ errors.
The logical error $\epsilon_l$ for exact codes is a step function of weight $t$.
This means that a weight-$(t+1)$ error can cause a logical error in the \emph{worst} case.
However, for quasi codes it is likely a monotonically increasing smooth function of it,
see Fig.~\ref{fig:dist}.
One has to introduce a cut-off $\epsilon_l^\wedge$ so that
a quasi code distance is $d_c=2t+1$ if $\epsilon_l(t)\leq \epsilon_l^\wedge$.
Here $\epsilon_l^\wedge$ is nothing but the accuracy parameter in a QEQ computation.
Furthermore,
as there are two types of quasi codes, weak and strong ones,
we can define weak and strong quasi code distances.
A quasi code distance is weak (strong) if the quasi code is weak (strong).
It is easy to see the weak distance refers to the average case,
while the strong distance refers to the worst case of physical errors making a logical error.

The differences between FTQ and QEQ computation we discussed above are 
summarized in Table~\ref{tab:FT_QEQ},
which also includes transversality that was studied previously~\cite{WZO+20}
and extended to cases with quasi encoding here (see Appendix~\ref{sec:trans_ae}). 
This is a justification of Fig.~\ref{fig:land},
and we shall recall that QEQ computation also shares some features with NQ computation as we discussed.

\begin{table*}[t!]
    \centering
    \setlength\extrarowheight{5pt}
    \begin{tabular}{|c|c|c|} \hline \Xhline{3\arrayrulewidth}
             & FTQ computation     & QEQ computation \\ \hline
    Code     & exact & quasi-exact      \\ \hline
    Accuracy & arbitrarily small    &  not arbitrarily small but tunable \\ \hline
    Algorithm & arbitrary accuracy as input & fixed accuracy as input or output \\ \hline
    Universality & SU($2^n$) & SU($2^n$) with gate-cells \\ \hline
    Code distance & $d_c=2t+1$ s.t. $\epsilon_l(t)=0$ & $d_c=2t+1$ s.t. $\epsilon_l(t)\leq \epsilon_l^\wedge$ \\ \hline
    Threshold  & $\epsilon_p\leq \epsilon_p^*$ s.t. $\epsilon_l\ra 0$ &  $\epsilon_p\leq \epsilon_p^*$
    s.t. $\epsilon_l\leq \epsilon_l^* $ \\ \hline
    Transversality & cannot be universal & can be quasi universal    \\ \hline
    Classification & strong & strong and weak \\ \hline \Xhline{3\arrayrulewidth}
    \end{tabular}
    \caption{Comparison between FTQ and QEQ computation
    in terms of accuracy,
    algorithm, universality, code distance, threshold, transversal logical gates,
    and classification of quasi codes.}
    \label{tab:FT_QEQ}
\end{table*}

\section{Valence-bond-solid codes}
\label{sec:VBSC}

\subsection{SU($d$) valence-bond solids}

Valence-bond solids~\cite{AKLT87} are the precursor of matrix-product states (MPS)~\cite{PVW+07,Sch11}
and also symmetry-protected topological (SPT) orders~\cite{CGW11,SPC11,CGL+12,CGL+13,DQ13a,DQ13b}.
It has been shown to be powerful for quantum computing in various contexts.
Here we treat them as quasi codes and provide a systematic study to reveal generic features of SPT order, quasi codes, and QEQ computation.

\begin{figure}[b!]
  \centering
  \includegraphics[width=.4\textwidth]{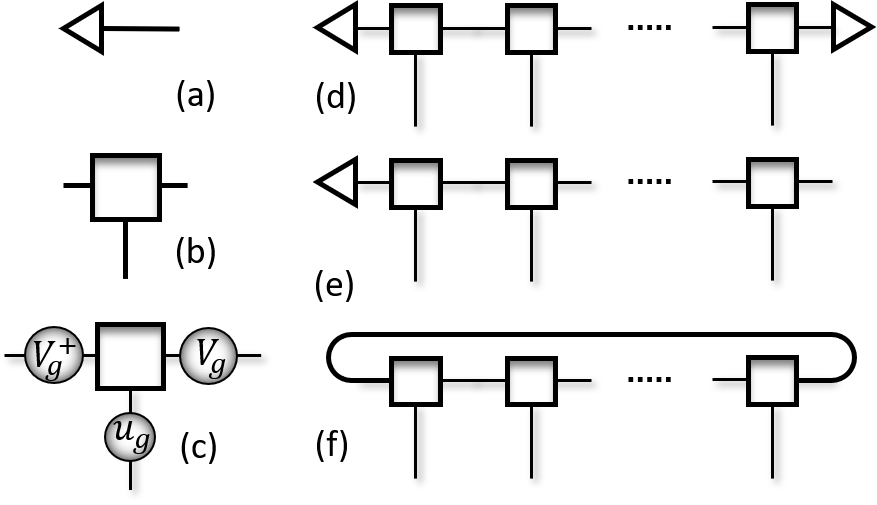}
  \caption{Matrix-product states. (a) A pure state $|\psi\ket$;
  (b) A three-leg tensor with the vertical one as the physical site;
  (c) The global symmetry on a tensor leaves it invariant;
  (d) MPS with OBC; (e) MPS with HBC; (f) MPS with PBC.}\label{fig:mps}
\end{figure}

An arbitrary finite-dimensional quantum state can be written as
\begin{equation}\label{}
  |\Psi\rangle=\sum_{i_1}^{d_1}\cdots\sum_{i_N}^{d_N} \T tr(B  A^{i_N}\cdots A^{i_1} )
  |i_1\rangle \cdots |i_N\rangle
\end{equation}
for $N$ local subsystems with local integer dimensions $d_n$,
and $B$ as the boundary operator,
which is $\I$ for periodic boundary condition (PBC),
$|l\rangle\langle r|$ for open boundary condition (OBC),
and $|l\rangle$ for the `half-boundary condition' (HBC) case.
The states $|r\rangle$ and $|l\rangle$ live in,
and the operators $A^{i_n}$ ($n=1,\dots,N$) act on a space of dimension $\chi$,
often known as the bond dimension.
For translational-invariant (TI) system,
we can always find site-independent $A^{i_n}$.
See Fig.~\ref{fig:mps}.

We consider 1D VBS with global SU($d$) symmetry.
The features our study reveal also apply to other types of VBS codes with different symmetries or in higher dimensions~\cite{AKLT87,GR07,KHK08,MUM+14,CFL+19}.
For its simplicity, we employ the
model with on-site adjoint irreps defined by
\begin{equation}\label{eq:H}
  H=\sum_{n=1}^N h_n + \frac{2}{3d} h_n^2:=\sum_{n=1}^N H_n
\end{equation}
with two-body local terms
$h_n:=\sum_{\alpha=1}^{d^2-1} T_n^\alpha \otimes  T^\alpha_{n+1}$,
with $\{T^\alpha\}$ the generators of SU($d$)
in the adjoint irrep for each site $n$.
Here `irrep' means irreducible representation.
The generators are $T^\alpha =-i\sum_{mn} f_{\alpha mn}|m\rangle\langle n|$
with structure constants $\{f_{\alpha mn}\}\in \mathbb{R}$ of SU($d$).

Now we study its ground states and primary excitations.
In the next section we will show how to use it as quasi codes.
In the VBS picture, each adjoint irrep is from a product of a fundamental irrep and its conjugate
since $\bf{\lambda}_d\otimes \overline{\bf{\lambda}}_{\bf{d}}=\bf{\lambda}_{d^2-1}\oplus \bf{\lambda}_1$,
for $\bf{\lambda}_r$ as an irrep with dimension $r$.
As $H_n$ is basically a projector,
the ground states can be exactly solved~\cite{WAR18}.

With PBC, the ground state is unique for $d=2$,
while doubly degenerate for $d>2$.
Denote the two ground states as $|G_L\rangle$ and $|G_R\rangle$.
The state $|G_L\rangle$ is specified by translation-invariant tensors
$A^i=\sqrt{\frac{2d}{d^2-1}} t^i$ for Gell-Mann matrices $t^i$
which satisfy $\T tr(t^it^j)=\frac{1}{2}\delta_{ij}$,
$[t^i,t^j]=if_{ijk}t^k$. 
The state $|G_R\rangle$ is defined with $(A^i)^*$, i.e.,
the system breaks a parity symmetry, denoted as $\mathbb{Z}_2^p$,
but preserves SU($d$) symmetry.
The parity refers to the interchange of a fundamental irrep $\bf{\lambda}_d$ and its conjugate $\overline{\bf{\lambda}}_{\bf{d}}$.
The generators for $\overline{\bf{\lambda}}_{\bf{d}}$ are $\{-(t^a)^*\}$.
For SU(2), the spin-$\frac{1}{2}$ irrep is the same as its conjugate,
hence the ground state is unique.
For SU($d$) with $d>2$, additional terms that explicitly break the parity symmetry can
split the degeneracy and pin one of the ground states as the unique ground state, denoted as $|G\ket$.
For OBC and HBC, we assume a unique bulk state is chosen, and w.l.o.g. we assume it is $|G_L\rangle$.

Observable are computed based on transfer matrix formalism.
The primary transfer matrix is
\begin{equation}
  \mathcal{M}:=\frac{2}{d}\sum_m t^m\otimes t^{m*},
\end{equation}
which is a real hermitian matrix.
The eigenvectors are
  $|v_0\rangle= \frac{1}{\sqrt{d}}\sum_i |ii\rangle:=|\omega\rangle$,
  $|v_a\rangle= \sqrt{2}\sum_{ij}t^a_{ij}|ij\rangle=\sqrt{2d} (t^a\otimes \I)|\omega\rangle$,
with $\langle v_b|v_a\rangle=\delta_{ab},\; \langle v_b|v_0\rangle=0$.
The eigenvalues are
$  \epsilon:=\frac{d^2-1}{d^2},\; \epsilon_1:=-\frac{1}{d^2}$
with
$\mathcal{M}|v_0\rangle=\epsilon|v_0\rangle,
  \mathcal{M} |v_a\rangle= \epsilon_1|v_a\rangle, \; a=1,\dots,d^2-1$.
For instance, to compute energy we first notice
the model is frustration-free,
but each local $h_n$ does not achieve its minimal value, which is $-d$ for a singlet as its eigenstate,
while in a ground state the nearest-neighbor two-body local state is not a singlet;
instead, the ground-state energy of $h_n$ is $E(h_n)=-\frac{d^3}{2(d^2-1)}$.
Meanwhile, the ground-state energy of $h_n^2$ is
$E(h_n^2)=\frac{d^2(d^2+2)}{4(d^2-1)}$, which is clearly not the square of $E(h_n)$.

\begin{figure}[b!]
  \includegraphics[width=.8\columnwidth]{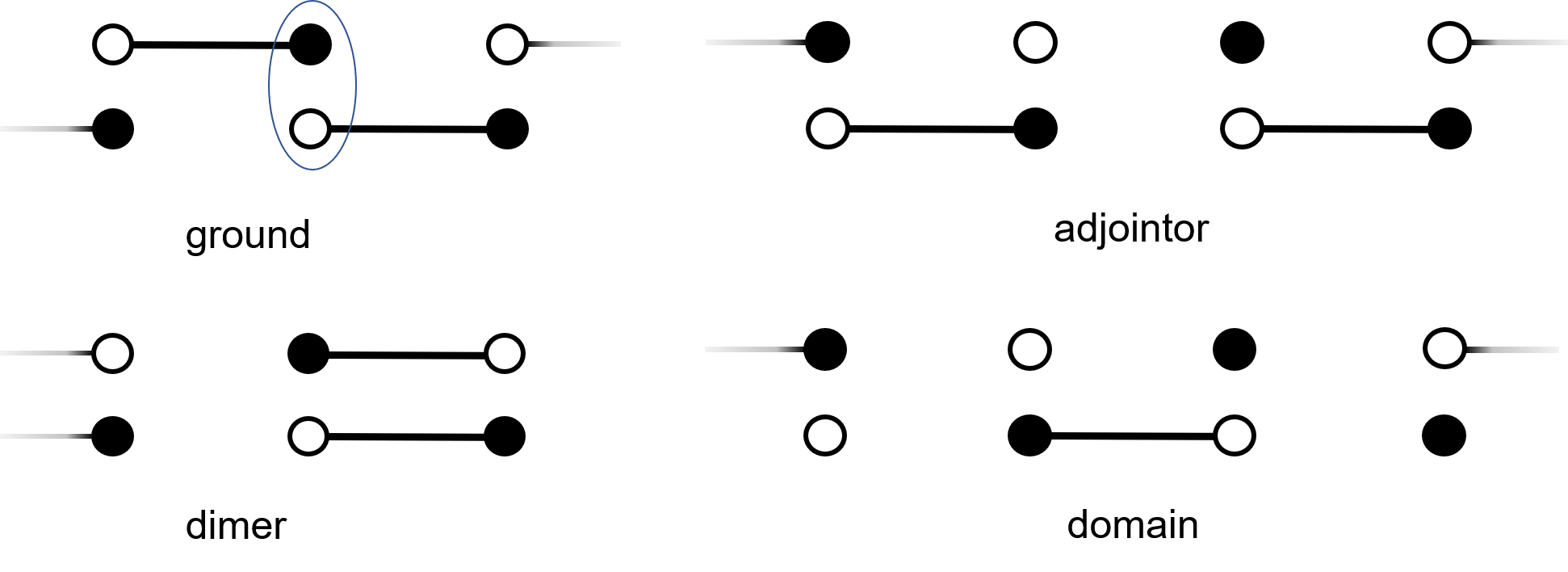}\vspace{-.5cm}
  \caption{Ground states and three types of approximate excitations.
  Solid lines represent singlet (bond).
  Each physical site is from a fundamental irrep (filled dot) and its conjugate (empty dot) in the same column.
  }\label{fig:vbcs}
\end{figure}

Excitations can be well described based on the single-mode approximation (SMA).
In the VBS picture, excitations are obtained by breaking bonds or rearranging them.
Some examples are shown in Fig.~\ref{fig:vbcs}.
Below we explain some terminology.
We call an excitation in the irrep $\bf{\lambda}_d$ as `soliton',
and its conjugate as anti-soliton.
When they live in between a ground state and an excited state,
they are called pseudo soliton (and anti-soliton).
An example of soliton is the well-known spinon which is a spin-$\frac{1}{2}$ excitation.
A broken bond is a single-mode which is a pair formed by a pseudo soliton and a pseudo anti-soliton,
also called an adjointor, as a generalization of `triplon'.
Solitons are confined, i.e., it takes energy to separate them apart,
and the region in between them could be dimerized.
A dimerized state is a product state as a sequence of no bond and double-bond.

When viewing excitations created by operators on the physical sites,
the single-modes are equivalent to actions of $T^a_n$ on any site $n$.
Correlation function decays exponentially with $\bra T^a_n T_m^b \ket=\frac{d^3}{2D}\chi^r \delta_{ab}$ for $\chi:=-1/D$,
$D:=d^2-1$, and $r=|n-m|$ as the separation.
Furthermore,
a unitary operator $U_g\in \text{SU}(d)$ acting on a site $n$ is equivalent to
a $V_g\in \text{SU}(d)$ conjugating on site $n$ due to the symmetry condition (see Fig.~\ref{fig:mps}(c)).
The modes appear in pair, so we term this type of excitations as `bi-mode'.
Similar with single-modes,
the bi-modes are not exact but good approximate excited states which will be used in our error-correction study.
For PBC there could be domains and domain walls which are solitons.
A pair of solitons is called a `vison' and a pair of anti-solitons is an anti-vison.
A domain with zero size, i.e., only the four solitons, is a bi-mode,
and a domain is created by drifting visons and flipping the bonds in between.
There is no confining force between a vision and an anti-vision so that
the size of a domain can change at will.




\subsection{SU($d$) VBS codes}
\label{sec:nVBSC}

Based on our knowledge of the model described above,
we can define VBS codes.
There are three types of encoding:
a) The bulk code: using bulk states ground-state degeneracy (GSD);
b) The edge code: using edge states;
c) The holographic code: using edge-bulk map $|l\rangle \mapsto |\Psi\rangle$ as encoding.
For bulk code,
the GSD of $H$ comes from the spontaneous symmetry-breaking (SSB) of a global symmetry.
For edge code, the code space comes from the edge operator $B=|l\ket\bra r|$,
and the degeneracy comes from the missing term $H_N$ near the edge.
For holographic code, the code space come from the `half-edge' operator $B=|l\ket$.
We can think of this by fixing the other edge by coupling to an additional system at site $N+1$,
and adding a new local term $H'_{N}$ to minimize the energy.

Different from stabilizer codes,
the VBS model is frustration-free, i.e., each local Hamiltonian term $H_n$ takes the minimal value,
while $[H_n,H_{n+1}]\neq 0$ $\forall n$.
The local terms $\{H_n\}$ can still be viewed as stabilizers,
or nullifiers in the setting of continuous-variable codes.
The local evolution $e^{-it H_n}$ with $t\in \mathbb{R}$ and their product has trivial effect on the code.
The local terms $\{H_n\}$ can be partitioned into two sets acting on even and odd sites,
$H_\text{even}$ and $H_\text{odd}$,
so that each set can be measured and the two sets are measured alternatively,
which slightly generalizes the case for stabilizer codes.

Besides quasi QEC,
we also study QEQ computation with the VBS codes.
Logical gates are unitary operators $U$ that commute with the model $H$ on the code space.
Namely, let $P$ be the projection onto the code space, then
\be P[U,H]P=0,\; [U,P]=0.\ee
Logical gates can be viewed as `emergent symmetry' which is a symmetry of $H$
only on the code subspace instead of $H$ itself.

Below we compare the three types of VBS codes, see Table~\ref{tab:vbscode}.
We study the encoding, low-weight errors, high-weight errors, code distance, threshold,
 logical gates, and preparation, etc.
We find that the edge and holographic codes behave similarly since they both depend on edge states
and the SSB of the global SU($d$) symmetry at the edge.
The bulk code preserves the SU($d$) symmetry while breaks the parity symmetry,
so it supports different logical gates from the other two codes.
Nevertheless, all three codes have similar quasi code distance and quasi threshold.
The quasi threshold is defined by introducing a dependence of distance on the system size,
instead of using other methods such as code concatenation.
This feature is relevant to the exponentially decay correlation function in the system.
Due to the SPT order, the exact code distance shall be a constant~\cite{ZZ14,GKS+19};
however, the weak quasi code distance can be larger and
we will show that VBS codes are useful for QEQ computation.

\begin{table*}[t!]
    \centering
    \setlength\extrarowheight{5pt}
    \begin{tabular}{|l|c|c|c|}\Xhline{3\arrayrulewidth}  \hline
             & Bulk code    &  Edge code   & Holo. code \\ \hline
Code space    & bulk SSB GSD  &  edge states & edge states \\ \hline
TLG          & $X$, $Z(\frac{2\pi}{d})$ & SU($d$) & SU($d$) \\ \hline
topo./global & $X$ global, $Z$ topo. & global & global \\ \hline
Code distance  & $d_c(X)=N$, strong; $d_c(Z)\in o(\sqrt{N})$, weak & $d_c\in o(\sqrt{N})$, weak   & $d_c\in o(\sqrt{N})$, weak\\ \hline
Threshold     & $\epsilon^*_p(X)=\frac{1}{2}$, $\epsilon^*_p(Z)\in \frac{o(\sqrt{N})}{N}$ & $\epsilon^*_p\in \frac{o(\sqrt{N})}{N}$ &
$\epsilon^*_p\in \frac{o(\sqrt{N})}{N}$ \\ \hline \Xhline{3\arrayrulewidth}
    \end{tabular}
    \caption{Comparison of the three primary types of VBS quasi codes: the holographic code,
    edge code, and bulk code in terms of code space, transversal logical gates,
    gates being topological or global, code distance, and threshold.}
    \label{tab:vbscode}
\end{table*}

\subsubsection{Holographic and edge codes}

In this section we study the holographic and edge VBS codes together since they behave similarly.
A primary error model for the holographic code was introduced earlier~\cite{WZO+20},
here we study the error model in more details.
For the holographic code,
the bulk effectively induces the encoding operation.
The encoding is explicit:
the original logical system is a part of the final physical system.
For the edge code, the encoding is implicit:
we still need an assignment between the logical states and the physical states.

\begin{figure}[b!]
  \centering
  \includegraphics[width=.7\columnwidth]{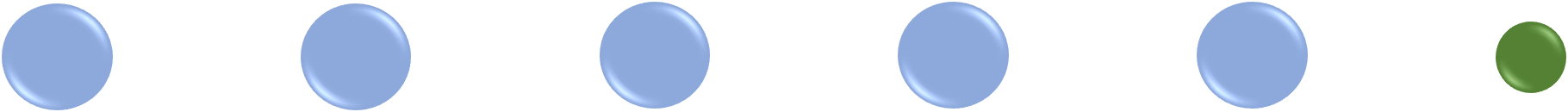}
  \caption{Schematics of a holographic VBS code.
  Big dots are the bulk sites, while the small dot is the edge site.
  A codeword is prepared by using a unitary operator $W$, as the dilation of the depolarizing channel $\mathcal{E}$,
  acting on the edge site with each bulk site sequentially, with the edge state initialized at a logical state $|\alpha\ket$ and each bulk site initialized at the state $|0\ket$.
  }
  \label{fig:holo}
\end{figure}

We first study the holographic code.
From the viewpoint of the edge,
it undergoes a sequence of channels $\mathcal{E}$ formed by $\{A^i\}$,
which is a depolarizing channel.
The logical state $|\alpha\ket$ is distributed over the whole system.
The channel $\mathcal{E}$ can be dilated to a unitary operator $W$ such that
$W |0\rangle=\sum_{i} A^i |i\rangle$, which is an isometry.
The encoding isometry $V$ is from the product of $W$. See Fig.~\ref{fig:holo}.


The reduced one-body bulk state $\rho^{(1)}$ is not the completely mixed state $\I$
due to the exponentially decaying bulk-edge correlations.
We label the edge as the ($N$+1)th site, and others from 1 to $N$ sequentially.
The edge before encoding can be viewed as the 0th site.
Let $\sigma_{n}:=\C E^{n-1}(|\alpha\ket\bra \alpha|)$ as the logical states of the edge, then
the local state $\rho_n^{(1)}$ ($n\in [1,N]$) of the $n$th bulk site is
\be  \rho_n^{(1)}=\sum_{i j}  \T tr [  \sigma_{n} A^jA^i  ] |i\ket\bra j|, \ee
which depends on $\sigma_{n}$.
Both $\sigma_{n}$ and $\rho_n^{(1)}$ converges to the completely mixed state exponentially as $n$ increases.
The final state of the edge can be obtained explicitly,
which is $\sigma_{N+1}= \I/d +
2\chi^N \sum_a t^a t^a_{\alpha\alpha}\ra \I/d$,
for 
$t^a_{\alpha\beta}:=\bra \alpha|t^a |\beta\ket$.
This means local state $\rho_n^{(1)}$ contains a bit information of the logical state $|\alpha\ket$,
while the set of all local states is needed to completely reconstruct $|\alpha\ket$.

It is convenient to describe the error model using the bond space,
and we will show later that it is equivalent to errors acting on physical sites.
In the bond space, we can treat a local error as the breaking of a bond.
Denote the bond or link $(n,n\pm 1)$ as $n\pm$.
For the local Gell-Mann error set $\{t_{n+}^a\}$, we find
\be \bra \psi_\alpha|  t_{n+}^a  |\psi_\beta\ket = \chi^n t^a_{\alpha\beta}, \label{eq:qde}\ee
and also
$\bra \psi_\alpha|  t_{n-}^a  |\psi_\beta\ket = \chi^{n-1} t^a_{\alpha\beta}$
for codewords $|\psi_\alpha\ket$ and $|\psi_\beta\ket$.

Consider the depolarizing channel $\C N_n$ on any site $n$ formed by
$\{E_k\}=\{E_0=\sqrt{1-p}\I, E_a=\sqrt{p}\sqrt{\frac{2d}{D}}t^a\}$ for $D:=d^2-1$
at each link.
The noise $\C N_n$ can be detected by the local term $H_n$.
From the quasi QEC theory, we find the recovery operators
$R_k=PE_k^\dagger /\sqrt{d_k}$ for $d_0=1-p$, $d_a=p/D$,
and $R_0=P$, $R_a=\sqrt{2d}Pt^a$.
We find
\be \hat{\sigma}_f=\C V^\dagger \C R \C N \C V (\hat{\sigma})= \hat{\sigma}+ \chi^{2n} 2d \sum_a t^a \hat{\sigma} t^a. \ee
This means that after the local recovery at link $n$,
there is a noisy map $\C Q_n$ which makes a logical error depending on $\chi^{2n}$.
For the QEC instrument,
we average over random errors that
occur on the system, then
\be \C V^\dagger \C Q \C V(\hat{\sigma}) =\bra \hat{\sigma}_f \ket_N= \hat{\sigma}+ \frac{2d}{N(D^2-1)} \sum_a t^a \hat{\sigma} t^a, \ee
which converges to $\hat{\sigma}$ in the large-$N$ or large-$d$ limit.
Also $\T tr (\C V^\dagger \C Q \C V(\hat{\sigma}))=1+\frac{D}{N(D^2-1)}\ra 1$ in the large-$N$ or large-$d$ limit,
which means $\C Q$ is aCPTP.
The trace distance on Choi state is 
\be D_t=\frac{D}{2N(D^2-1)}\approx \frac{1}{2ND}, \ee
which sets the accuracy of the code.

The quasi QEC above also applies to bounded operators in the span of
the local Gell-Mann error set $\{t_{n+}^a\}$ together with $t^0\equiv \I$, hence it has the LKS property.
Say, for $V_{n+}=\sum_a v_a t^a_{n+}$, we find
$\bra \psi_\alpha|  V_{n+}  |\psi_\beta\ket =
v_0 \delta_{\alpha\beta}+\chi^n \sum_k v_k t^k_{\alpha\beta}$.
The uncorrectable part decays with $\chi$,
which after average over the system agrees with the analysis above.

The local error on a single bond is single-mode and
it is equivalent to the bi-mode model and is simpler.
First, the action of a local operator $T^a$,
as the adjoint representation of $t^a$, is
$T^a\sum_i A^i |i\ket=\sum_i [A^i,t^a] |i\ket$,
which means it is a superposition of
$t_{n+}^a$ and $t_{n-}^a$ acting on the bond space at site $n$.
For bi-mode error caused by local unitary in the symmetry,
the virtual action is a conjugation by a unitary $U\in \text{SU}(d)$.
For $U=u_0\I+\sum_a u_a t^a$,
it is clear that it is quasi detectable with correction in the order $\chi^n$.
It can be detected by $H_n$ and $H_{n-1}$ and quasi corrected by $\C R_n$ and $\C R_{n-1}$
since $U$ is in the Kraus-span of $\C N_n$.
Therefore, it is equivalent to the single-mode local error model.


We have assumed that the separation between pseudo solitons is minimal.
When they are separated by a dimerized region,
the syndrome can also detect them.
A recovery scheme is needed which can reduce the size of dimerized region,
and bring pseudo soliton pairs together, and then annihilate them.
This is analogue with the minimal-weight perfect-matching algorithm that is commonly used for toric code~\cite{DKL+02}.
As the system is 1D, the matching of soliton pairs can be rather easily done.
This will be discussed in more details below for the bulk code.

For higher-weight errors, there are further uncorrectable part from the bulk correlations
and the accuracy (i.e., $D_t$) will get worse.
Local errors are detected individually and corrected also locally.
Here we provide a study when the weight $t$ is much smaller than the system size $N$.
For two-site correlation functions we find
\be \bra \psi_\alpha|  t_{m+}^a t_{n+}^b  |\psi_\beta\ket = \chi^{n-m} \delta_{ab}\delta_{\alpha\beta}/2d+\chi^nh_{bac} t^c_{\alpha\beta}/2, \label{eq:qco} \ee
for $n>m\geq 0$, $h_{bac}=d_{bac}+if_{bac}$,
and $d_{bac}$ are also structure constants of SU($d$).
For two unitary errors  $V_{m+}=\sum_a v_a t^a_{m+}$,
$W_{n+}=\sum_b w_b t^b_{n+}$, we find
\be \bra \psi_\alpha|  V_{m+} W_{n+}  |\psi_\beta\ket =
f_0 \delta_{\alpha\beta}+\sum_k f_k t^k_{\alpha\beta}, \ee
for $f_0:=v_0w_0+\chi^{n-m}\sum_{k\neq 0} v_kw_k/2d$,
$f_k:= \chi^n v_0w_k+\chi^m w_0 v_k + \chi^n \sum_{bc} w_b v_c h_{bca}/2$.
For weight $t=2$ we find the recovered logical state is
\be \hat{\sigma}_f = (1+D\chi^{2r})\hat{\sigma}+ 2d(\chi^{2n}-\chi^{2m}) \C E(\hat{\sigma}), \ee
for $\C E$ as the map formed by $\{t^a\}$,
$r$ as the separation between two errors at $n$ and $m$.
We see that there are exponentially small terms due to the bulk separation $r$
and the separations $n$ and $m$ from the edge.
The minus sign for $\chi^{2m}$ is an interference effect,
which may also be constructive for higher-weight errors.

In general, there are edge terms and bulk terms in $\hat{\sigma}_f$,
which both decay exponentially with separations.
For weight $t$, there are $t$ edge terms and $\frac{t(t+1)}{2}-t$ bulk terms.
We shall take the average
\bea &&\sum_{r=1}^{N-t+1} \frac{\chi^{2r}}{N-t+1} \approx \frac{1}{(N-t+1)D^2},\; \\ \nonumber
&&\sum_{r=1}^{N-t+1} \chi^{2r} \frac{(N-t+2-r)}{\binom{N-t+2}{2}} \approx \frac{2}{(N-t+2)D^2},\eea
when $\chi^{2(N-t)}\ra 0$.
We find the trace distance
\be D_t \approx \frac{t^2}{2D(N-t)},\ee
which reproduces the $t=1$ case.
The $t^2$ factor is due to the bulk effect as the number of bulk separations is $O(t^2)$,
and $(N-t)$ is the effective system size when there are $t$ errors.

We see that the accuracy of the code against weight-$t$ error, measured by $D_t$,
is a smooth function of $t$ (also see Fig.~\ref{fig:dist}).
We shall introduce a cut-off $\epsilon^\wedge_l$ which then defines a code distance,
or vice versa.
This is distinct from the case of exact codes.

We can then determine the quasi threshold.
Consider local noise model as a depolarizing channel $\C N$ with $p\in[0,1]$ as the physical error rate.
If the distance is $d_c=2t+1$,
we consider the probability of up to $t$ errors $p(t)$ as the success rate,
and $1-p(t):=\epsilon_l$ as the logical error rate.
Similar with repetition code,
$p(t)=\sum_{x=0}^t \binom{N}{x} p^x (1-p)^{N-x}$.
From Chernoff bound, we can determine the quasi threshold as
\be \epsilon_p^*= \frac{t}{N}, \; t\in o(\sqrt{N}),\ee
and the logical threshold, which is the accuracy of the code, is $\epsilon_l^*=\frac{t^2}{2D(N-t)}$.
In order to ensure $\epsilon_l^*\ra 0$, we shall choose $t \in o(\sqrt{N})$,
which leads to $\epsilon_p^*\ra 0$.
This means when the code becomes exact by $N\ra \infty$,
the code distance grows no faster than $\sqrt{N}$,
the threshold $\epsilon_p^*$ power-law decays to zero.
For finite $N$, which is physically the case in practice,
there is a finite threshold $\epsilon_p^*$.
From a thermodynamic point of view,
the average number of errors $t$ corresponds to the temperature, $\beta_\text{th}$,
with $t\sim N e^{-\Delta\beta_\text{th}}$ for $\Delta$ as the gap of a VBS model.
A precise estimation of their relation will depend on more details of a system.

Logical gates follow from the global symmetry of $\mathcal{E}$ with
\begin{equation}\label{eq:symcond}
  \sum_j u_{ij}(g) A^j=\breve{U}(g) A^i \breve{U}(g)^\dagger
\end{equation}
for $U(g)=[u_{ij}](g)$ of size $d^2-1$, $\breve{U}(g)$ of size $d$,
for $g\in \text{SU}(d)$~\cite{CGW11,SPC11}.
Now it is clear that the code is quasi universal and transversal for SU($d$) with
\be V \breve{U}(g)|\alpha\ket= U(g)\otimes U(g)\otimes \cdots \otimes U(g) \otimes \breve{U}(g)  |\psi_\alpha\rangle, \ee
for $N$ factors of $U(g)$ acting on the bulk and $\breve{U}(g)$ acting on the edge, for $g\in \text{SU}(d)$,
and $V$ as the encoding isometry.

The edge code behaves similarly with the holographic code.
Here we highlight their differences.
Due to the bulk-edge correlation,
the bulk local states will depend on the edge states.
The local state $\rho^{(1)}_n$ is
\be \rho^{(1)}_n=\sum_{i_n,j_n} \T tr((B\otimes B^*)\C M^{N-n} (A^{j_n}\otimes (A^{i_n})^*) \C M^{n-1}) |j_n\ket\bra i_n|. \ee
If the bulk site $n$ is close to the edge $1$ or $N$,
the correlation between bulk and edge is apparent.
If $n\gg1$, we find
\be \rho^{(1)}_n=\frac{1}{\epsilon d^3}\I +\chi^{N-n} \frac{1}{\epsilon d^2}\sum_{ijk}h_{jik} t^k_{rr}|j\ket\bra i|. \ee
Also we find similar form for $n\ll N$.
We see that $\rho^{(1)}_n$ converges exponentially to $\I$.
Deep in the bulk, $\rho^{(1)}_n$ is almost $\I$.
The error model is the same with the holographic code.
To perform logical gates, observe that
a global symmetry $\otimes_n U_n$ acts as $(\breve{U}\otimes \breve{U}^\dagger)|l,r\ket$.
No entangling gates can be induced on $|l,r\ket$ by global symmetry operations.
This means effectively one edge is enough for the computation.
As a result, we can use $|l\ket$ as the logical states.
If the system is large enough, the effect of the other edge $|r\ket$ can be omitted,
and this is essentially the same as the holographic code.

For the two edge codes, initialization and readout can be done by coupling the edge to external ancilla.
The system shall be cooled to a low temperature comparable to the gap to suppress excitations,
and active error correction can also be employed.
For the holographic code, one edge $\overline{\bf{\lambda}}_{\bf{d}}$ already is coupled to an irrep $\bf{\lambda}_d$
to form a singlet.
This is the ground state of the additional term $H_{N+1}'$ between the edge site $N$ and the ancilla at site $N+1$.
The term $H_{N+1}'$ can be constructed as a projector $P_{(\bf{\lambda}_d)}$,
which is a projector onto the parts except the irrep $\bf{\lambda}_d$,
which appears in the tensor-product $\bf{\lambda}_d \otimes \bf{\lambda}_{d^2-1}$,
similar with bulk terms.
We use an ancilla $\overline{\bf{\lambda}}_{\bf{d}}$ to couple the other edge at site 1.
To initialize it, one can measure the ancilla and project it onto a state $|l\ket$,
which then initializes the code at logical state $|\psi_l\ket$ depending on $|l\ket$.
The code distributes the information of state $|l\ket$ into the global state $|\psi_l\ket$.
It is similar to initialize the edge code, wherein one need to project out the two ancillas.
After computation by global symmetry operations,
the logical state $|f\ket$ can be determined by the values $\bra f|t^a|f\ket:=t^a_f$ which is
\be t^a_f= \bra \psi_f|t_{N+1}^a  |\psi_f\ket+\sum_n \bra \psi_f|  T_n^a  |\psi_f\ket. \ee
That is to say, it is readout from the local values of $T_n^a$ for the edge code,
and also $t_{N+1}^a$ for the holographic code.
The values of $T_n^a$ may not be easy to measure precisely since some of them may be very close to zero.
The final state may also be a mixed state.
We find a different method by effectively changing the boundary condition as follows.
Perform a projection on the other ancilla onto a state $|g\ket$,
which induces a boundary state $|r\ket$ and the final state is a MPS with boundary operator $B=|f\ket\bra g|$.
We can measure the value of the local term $h$ between the first site and the final site which is
\be E(h)= \frac{d^3}{2D^2}-\frac{d^4}{2D^2}|\bra g|f\ket|^2 \ee
and deduce the value $|\bra g|f\ket|^2$.
Now if we can vary $|g\ket$ over an informationally complete set,
then the logical state can be easily obtained 
following from state tomography.
For any $d$, such a set is believed to exist~\cite{RBS04},
which has $d^2$ projectors that are not orthogonal to each other, in general.

\subsubsection{Bulk code}

In this section, we study quasi QEC of the bulk VBS code.
This has been studied in Ref.~\cite{WAR18,Wang20a},
here we estimate the threshold and compare with the other two types of codes.
The two ground states are not exactly orthogonal
as $\bra G_L|G_R\rangle\propto (\frac{1}{d-1})^N$, which vanishes only for large $d$ or large $N$.
This means the encoding for bulk code $|0\ket:=|G_L\rangle$, $|1\ket:=|G_R\rangle$ is quasi exact.
We will omit this effect by considering the large-$N$ limit.

Different from the edge code and holographic code,
the global symmetry SU($d$) is logically the identity gate.
For any $d>3$, the bulk code is a single logical qubit.
The logical $X_L$ gate is the generator of the broken parity symmetry,
which can be realized transversally as $X_L=\vec{\Pi}=\otimes_n \Pi_n$ for $\Pi$ as a permutation,
which exchanges the irrep $\bf{\lambda}_d$ and its conjugate $\overline{\bf{\lambda}}_{\bf{d}}$.
The preserved global symmetry induces a logical $Z$-rotation gate $e^{i\frac{2\pi}{d}Z}$,
which is a twist operator, $F=\otimes_n e^{i\frac{2\pi}{N}\hat{O}_n}$,
for $\hat{O}$ as an operator in the Cartan subalgebra of SU($d$)~\cite{WAR18}.
The operator $F$ is a diagonal matrix and becomes complex-conjugated after $\vec{\Pi}$ since
\be \vec{\Pi}F\vec{\Pi} = F^\dagger. \ee
The code can be viewed as a generalization of the Ising model with the anti-commutation relation
\be \{\Pi,T^a\}=0.\ee
The local operator $\Pi$ acts like Pauli $X$ operator, and $T^a$ acts like Pauli $Z$ operator.
The term $h_n$ is analog to the Ising interaction, except that it is augmented with the SU($d$) symmetry,
due to which the generators $T^a$ are equivalent with each other.
It is proper to treat the bit-flip gate $X_L$ as acting on the physical site,
while the phase-flip gate $Z$-rotation acting on the virtual bonds, i.e., the bond space.

To establish an error model, we introduce $X$-type errors caused by $\Pi_n$ flipping bond directions
and $Z$-type errors caused by operators breaking bonds due to the SU($d$) symmetry.
First, we discuss $X$-type errors and $Z$-types errors separately
since they arise from different mechanism,
and later on we discuss $Y$-type errors.
The scenario for $X$-type errors is similar with Ising model,
which also has SSB of a global symmetry.
The local terms $h_n$ anti-commutes with local errors
\be \Pi_n h_n \Pi_n=-h_n,\; \Pi_{n+1} h_n \Pi_{n+1}=-h_n, \ee
while
$\Pi_n\otimes \Pi_{n+1}$ will create a direction-flipped bond,
together with a vison-antivison pair,
and $[\Pi_n\otimes \Pi_{n+1}, h_n]=0$.
A single error $\Pi_n$ can be detected by $h_n$ and $h_{n-1}$.
By measuring all $h_n$ and then counting the global parity,
the locations of $\Pi_n$ can be deduced and then applying $\Pi_n$ can correct the errors.
Furthermore, it is also clear that the code distance $d_c(X)=N$,
and the threshold for local $\Pi_n$ errors
is $\frac{1}{2}$, the same as the repetition code.

The effects of $Z$-type errors are similar with that for the edge codes
except that there are only bulk states.
Due to the exponentially decaying bulk correlations,
the local states of bulk sites can be different for different codewords,
hence the code is a quasi code.
Only one-body local states $\rho_n^{(1)}$ are the completely mixed state,
so local error $\C N_n$ breaking the bond at link $n$
is exactly correctable with the same recovery scheme $\C R_n$
as for the edge codes.
Note for arbitrary superposition $|\psi\ket=a|G_L\ket+b|G_R\ket$,
the bond dimension is $2d$ and the local tensors are block diagonal,
with each block for a ground state.
On a link $n$, an error acts as $\C N_n$ on one block,
and $\C N_n^t$ on the other block.
Similar with edge codes,
for higher-weight errors there are uncorrectable bulk terms which scale with their spatial separations.
For $t$ errors, there are $t(t-1)/2$ bulk terms.
The trace distance $D_t$ is $\frac{t(t-1)}{2D(N-t)}$,
and the quasi threshold for $Z$-type errors will be the same with edge codes.

Using the notion of soliton,
the $X$-type errors introduce solitons and domains,
while the $Z$-type errors introduce pseudo solitons.
Now for $Y$-type errors,
there are both solitons and pseudo solitons, and domains.
To model $Y$-type errors,
we consider on-site discrete error set $\{\Pi^i(T^{a})^j\}$ for $i=0,1,j=0,1,2$.
The operators $(T^{a})^2$ can be viewed as a second-order effect of $T^a$,
which is further equivalent to the set $\{t^a\}$ acting on the bond space.
We find the error correction procedure contains two steps:
\begin{enumerate}
  \item First, identify the domains and then use error correction to delete the wrong domains;
this annihilates all solitons and corrects $X$-type errors;
  \item Second, identify broken bonds and use error correction to annihilate all pseudo solitons;
this corrects $Z$-type errors.
\end{enumerate}
There could also be dimerized regions between pseudo solitons,
and they can be easily corrected and all pseudo solitons are annihilated at the end.
We note that the details of the procedure above will highly depend on practical implementations,
which we do not describe here.

The code distance for $X$-type errors is the system size $N$,
similar with 1D Ising repetition code.
As a domain wall can drifted away randomly without causing a net energy,
it is not self-correcting for $X$-type errors.
The code distance for $Z$-type errors, which is weak and quasi-exact,
is the same as edge codes, $d_c(Z)\in o(\sqrt{N})$,
also for $Y$-type errors.
As a result,
the quasi threshold for $Y$-type errors shall be similar with $Z$-type errors. See Table~\ref{tab:vbscode}.
In practical systems,
the $X$-type errors and $Z$-type errors might be highly biased.
Breaking a bond at link $n$ will cause energy penalty for one term $h_n$,
while flipping a bond at link $n$ will cause energy penalty for two terms $h_{n-1}$ and $h_{n+1}$,
so it seems it is easier for $Z$-type errors to occur.
However, a single broken bond caused by $\C N_n$ will only lead to a leakage out of the code space.
It requires at least two broken bonds to yield an uncorrectable logical error,
which will then cause similar energy penalty with flipping a bond.
Therefore, it would be desirable to develop methods to suppress $Z$-type errors in practice.
Finally, we note that
a transversal and quasi universal scheme based on the bulk VBS code is unknown yet.

\section{Discussion and Conclusion}
\label{sec:disc}

In this work we studied the framework of quasi-exact quantum (QEQ) computation,
which is computation based on quasi codes.
Here we discuss further on several points.

The tuple form 
$\bra \C V, \C N, \C R, \vec{\lambda}, \epsilon \ket$
allows optimization schemes for finding good quasi codes.
This can be done for codewords expressed as a manifold of parameterized family of matrix-product states or quantum circuits, and optimize the accuracy $\epsilon$ while keeping the error model and recovery scheme fixed.  
This is distinct from the optimization for usual approximate codes~\cite{BO10} which search for recovery channels while keeping the error model and encoding fixed. 
More generally, the idea of using scaling parameters agrees with the technique of optimization applied for various coding tasks~\cite{BF20,BL20}, applications of JNR theory~\cite{buoni1978joint,xie2020observing,spitkovsky2018signatures,chen2016geometry}, quantum control, machine learning, etc. 

The valence-bond-solid (VBS) codes are important classes of quasi codes.
Different from all exact codes, we establish their code distances to be quasi-exact and of weak type.
Therefore, they show some novel features that are not observed for exact codes including stabilizer codes.
In particular, the SU($d$) VBS edge or holographic codes
are transversal and quasi universal for SU($d$).
However, the QEQ computation supported by quasi codes is slightly weaker than
the standard fault-tolerant quantum (FTQ) computation.
The continuous symmetry of VBS codes is one of the novel features on one hand,
yet on the other hand, the symmetry requires delicate scheme to avoid being broken down to discrete symmetry or careful control to execute gates.

In general,
the VBS codes we constructed can be extended to quantum phases of matter,
if we view a VBS state as a special point of a certain SPT phase.
The power of SPT phases has been explored in the setting of measurement-based quantum computation~\cite{SWP+17},
which shows that if special states of a SPT phase is powerful, namely,
can be used to execute a non-Abelian Lie group,
then most likely all states in the same phase can also do so.
As discussed in Sec.~\ref{subsec:ex} of examples, 
concrete study of how quasi QEC works across various types of gapped phases of matter shall be investigated.



We also would like to add some comments on quasi codes
which could be transversal and quasi universal for SU($2^n$), for any $n$,
and in particular, codes with discrete gate set.
The codewords for such desired golden codes may support extensive entanglement
so that transversal operations on it, which do not change entanglement, can be quasi universal
and cannot be efficiently simulated on classical computers.
The (exact or approximate) symmetry group of such a code could be extensive, 
e.g., may be a higher-form symmetry~\cite{KS14,GKS+15},
so that the spectrum is rich enough to encode as many qubits as possible.
It would be important to see if there are candidate models of these kinds as quasi codes.

\section{Acknowledgement}
This work has been funded by
the Government of Ontario
and the Government of Canada through ISED (D.-S.W., R.L.),
the Startup program at ITP (D.W.),
and the National Science Foundation of China under Grants 61771377 and 61502376 (Y.-J.W.). 
Conversations with S.-J. Qin, T. Lan, C. Okay,
Y. Yang, G. Zhu are acknowledged.


\appendix

\section{Completely positive maps}
\label{sec:cp}

\subsection{Quantum channels and instruments}
\label{subsec:qchannel}

A completely-positive (CP) map $\Phi: \C D(\C H_1) \ra \C D(\C H_2)$ takes the form
\be \Phi(\rho)=\sum_i K_i \rho K_i^\dagger, \forall \rho \in \C D(\C H_1), \ee
and $K_i$ are often known as Kraus operators.
The minimal number of $K_i$ to represent $\Phi$ is called the rank of the map, rk($\Phi$).

When $\T tr(\Phi(\rho))=1$, i.e., $\Phi^\dagger(\I_1)=\sum_i K_i^\dagger K_i=\I_2$,
the map is also trace-preserving (TP).
Here, $\I_1\in \C P(\C H_1)$, $\I_2\in \C P(\C H_2)$.
A CPTP map is also called a quantum channel.
As we can see, quantum channels can change dimensions of Hilbert spaces.
For instance, isometric operators that are often used as encoding operators are channels.


A quantum instrument is a set of CP maps so that the sum of them is a quantum channel~\cite{DL70}.
Namely, an instrument can be denoted as
\be \C Q=\{\C Q_x\}_{x\in X}, \ee
for $\C Q_x: \C D(\C H_1) \ra \C D(\C H_2)$
and $X$ as the index set,
which is assumed to be discrete here.
We often treat $\C Q$ itself as the quantum channel it represents.
The main difference between channels and instruments that we refer to
is that there exists a classical index that labels each CP maps.
An instrument is also known as a `selective' channel in the same sense as we explained.

\begin{figure}
  \centering
  \includegraphics[width=.25\textwidth]{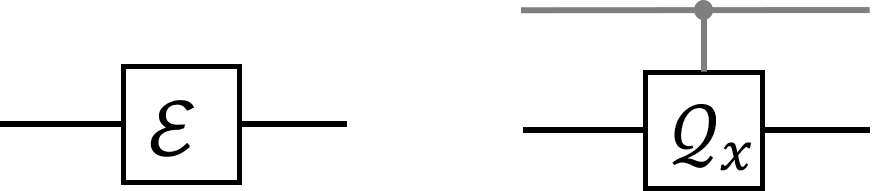}
  \caption{Quantum channel (Left), usually denoted as $\C E$,
  and quantum instrument (Right), denoted as $\C Q_x$ with a classical index $x\in X$.
  The gray wire is a classical control to record the index $x$.
}\label{fig:instr}
\end{figure}

The notion of instrument is broad as it describes many operations as special cases~\cite{CLM+14}.
For example,
a measurement, described as a positive operator-valued measure (POVM) $\{F_i\}$ with $\sum_i F_i=\I$,
is an instrument with each $F_i$ realized by a CP map $\C F_i$.
The Bell measurement employed in quantum teleportation is an instrument
since each outcome is recorded and feed forwarded for the Pauli byproduct correction.
A channel $\C E$ itself is an instrument.
A mixing of channels, $\sum_i p_i \C E_i$, is an instrument with each CP map as $p_i \C E_i$,
i.e., a channel $\C E_i$ but with a probability factor $p_i$.

\subsection{Distance measures}
\label{subsec:dist}

Here we survey some distance measures on CP maps.
The trace distance defined on $\C B(\C H)$ is
\be D_t(A,B)=\frac{1}{2}\|A-B\|_1= \frac{1}{2}\T tr\sqrt{(A-B)^\dagger(A-B)},\ee
$\forall A, B \in \C B(\C H)$. Here $\|\cdot\|_1$ denotes the trace norm.
The fidelity defined on $\C P(\C H)$ is
\be F(A,B)=\|\sqrt{A}\sqrt{B}\|_1^2, \; \forall A, B \in \C P(\C H).\ee
For quantum states, it is well known that
\be 1-\sqrt{F(\rho,\sigma)} \leq D_t(\rho,\sigma) \leq \sqrt{1-F(\rho,\sigma)},\ee
$\forall \rho, \sigma \in \C D(\C H).$ When one state is pure, say $\sigma$, the lower bound becomes $1-F(\rho,\sigma)$.
A single measure may not be enough to provide a precise measure of distance for some tasks,
instead, both the distance and fidelity can be employed together.

Similar with the case of states, there are two primary types of distance measures for CP maps based on
trace distance and fidelity.
For $\Phi, \Psi: \C D(\C H_1) \ra \C D(\C H_2)$,
the diamond-norm distance~\cite{KSV02} between them is
\bea D_\diamond(\Phi, \Psi) &=&\|\Phi \otimes \I- \Psi \otimes \I \|_{1\ra 1} \\ \nonumber
&=&\sup_\rho \|\Phi \otimes \I(\rho)- \Psi \otimes \I(\rho))\|_1,\eea
$\forall \rho \in \C D(\C H_1\otimes \C H'_1)$,
for $\I$ acting on an ancillary space $\C H'_1$ with dimension at least $d_1:=\text{dim}\C H_1$.

Any CP map $\Phi: \C D(\C H_1) \ra \C D(\C H_2)$ can be represented by the Choi operator
\be \omega_\Phi:= \Phi \otimes \I (\omega) \ee
for $|\omega\ket=\frac{1}{\sqrt{d_1}} \sum_{i=0}^{d_1-1} |ii\ket$ as a maximally entangled state,
$\omega=|\omega\ket\bra\omega|$.
The CP condition of $\Phi$ is equivalent to the positivity of $\omega_\Phi$.
The Kraus operator-sum representation of $\Phi$ can be found from the eigen-decomposition of $\omega_\Phi$,
and rk($\Phi$)=rk($\omega_\Phi$).

The trace distance between Choi states $D_t(\omega_\Phi, \omega_\Psi)$
is equivalent to the diamond-norm distance by~\cite{Wat18}
\be \frac{1}{2d_1}D_\diamond( \Phi, \Psi ) \leq D_t(\omega_\Phi, \omega_\Psi) \leq \frac{1}{2} D_\diamond( \Phi, \Psi ). \ee

The average entanglement fidelity~\cite{Sch96} between $\Phi$ and $\Psi$ is
\be F_E(\Phi, \Psi)=  F(\Phi \otimes \I (\omega),  \Psi \otimes \I (\omega)), \ee
while the worst-case entanglement fidelity is
$F_E'(\Phi, \Psi)= \min_{\psi_\rho} F( \Phi \otimes \I (\psi_\rho),  \Psi \otimes \I (\psi_\rho))$,
for $|\psi_\rho\ket$ as a purification of $\rho\in \C D(\C H_1)$.

The average gate fidelity is
\be F_G(\Phi, \Psi)= \int d\psi  F(\Phi(\psi), \Psi(\psi)), \ee
while the worst-case gate fidelity is
$F_G'(\Phi, \Psi)= \min_\psi  F(\Phi(\psi), \Psi(\psi))$,
for $|\psi\ket\in \C H_1$.
A notable connection between average entanglement fidelity and gate fidelity is~\cite{HHH99}
\be F_G=\frac{F_E d_1+1}{d_1+1}. \ee
Recall that $d_1$ is the dimension of $\C H_1$.
The quantity $F_E$ is the fidelity between Choi operators.
For two channels $\C E$ and $\C F$, it holds
\be 1-\sqrt{F_E(\C E,\C F)} \leq D_t(\omega_{\C E},\omega_{\C F}) \leq \sqrt{1-F_E(\C E,\C F)}.\ee
If one channel is unitary, $\C U$, then
\be 1-F_E(\C E,\C U) \leq D_t(\omega_{\C E},\omega_{\C U}) \leq \sqrt{1-F_E(\C E,\C U)}.\ee
In this work, we will employ the average-case fidelity measures and the diamond-norm distance.
In addition, we note that in the setting of quantum communication
there are also other quantities such as channel capacities~\cite{Llo97,DS05,BL20}.



\subsection{Subspace CPTP maps}
\label{subsec:scptp}

We now introduce a weaker type of channels suitable for QEC.
We only consider dimension-preserving CP maps.
\begin{definition}
  [sCPTP] A CP map $\Phi$ on $\C D(\C H)$ is subspace CPTP (sCPTP) if it is CPTP on a subspace $\C C \subseteq \C D(\C H)$.
\end{definition}
Let $P$ be the projector on $\C C$.
The condition for being sCPTP is as follows.
\begin{prop}
  A CP map $\Phi$ on $\C D(\C H)$ is sCPTP on $\C C \subseteq \C D(\C H)$ iff $\Phi^\dagger(P)=P$ and $\C P \Phi \C P= \Phi \C P$.
\end{prop}
\begin{proof}
  (Sufficiency.) Given the conditions, we need to show $\T tr(\Phi(\sigma))=\T tr\sigma=1$, $\forall \sigma \in \C C$.
  For any $\rho \in \C D(\C H)$, let $\sigma=P\rho P/p\in \C C$, for $p=\T tr(P\rho)$.
  Then $p\Phi(\sigma)=\Phi \C P (\rho)=\C P\Phi \C P (\rho)$,
  and we find $\T tr(\Phi(\sigma))=\T tr(\Phi^\dagger(P)\sigma)=\T tr\sigma=1$.

  (Necessity.) If $\Phi(\sigma)\in \C C$, $\forall \sigma \in \C C$, then
  $\T tr(\Phi(\sigma))=\T tr(\Phi^\dagger(P)\sigma)$, $\forall \sigma \in \C C$,
  which implies $\Phi^\dagger(P)=P$.
  Also it is easy to see $\C P \Phi \C P= \Phi \C P$ holds.
\end{proof}

Denote the partition of the whole space as $\C D(\C H)=\C C \oplus \C C^\bot$.
It is not hard to show that
a map $\Phi$ is sCPTP on $\C C$ iff its Kraus operators can be written as block-diagonal matrices simultaneously.
This can be proved by writing each Kraus operators as $K_i=\begin{pmatrix}A_i & C_i \\ D_i & B_i \end{pmatrix}$,
and then the condition $\Phi^\dagger(P)=P$ implies all $C_i=0$ and $\C P \Phi \C P= \Phi \C P$ implies all $D_i=0$.
Note that $A_i$ do not have to be unitary.
The matrices $D_i$ represent the leakage out of $\C C$,
while $C_i$ represent the leakage into $\C C$.
This basically means that there is no exchange of information between $\C C$ and $\C C^\bot$.
If a map $\Phi$ is sCPTP on $\C C$,
we say $\Phi$ is a logical operation on $\C C$.
Note that $\Phi$ is logical on $\C C$ does not in general imply $\Phi^\dagger$ is also logical.

There are some interesting special cases.
1) When $\C C$ is the whole space, the condition above reduces to the TP condition $\Phi^\dagger(\I)=\I$.
2) For any bounded operator $A\in \C B(\C H)$,
it is logical iff $A^\dagger$ is logical.
In block-diagonal form, the block of $A$ acting on $\C C$ is unitary.
3) A unitary operator $U$ is logical on $\C C$ iff $[U,P]=0$.

The property of being sCPTP is preserved under mixing and composition.
Namely, for several sCPTP maps $\{\C E_i\}$ on $\C C$,
the map $\sum_i p_i \C E_i$ with $\sum_i p_i=1$, $p_i\in (0,1)$ is also sCPTP on $\C C$.
The composition $\prod_i \C E_i$ is clearly also sCPTP on $\C C$.

From Stinespring dilation, a quantum channel can be realized by a unitary operator.
This can be extended to sCPTP maps for input states restricted to $\C C$.
Given a set of Kraus operators $\{K_i\}$ for a sCPTP map $\Phi$,
one can find $K'$ to make $\{K_i,K'\}$ as a CPTP map, $\Phi'$.
It is easy to see $K'$ has null effect on $\C C$.
Now one can find a dilated unitary $U$ for $\Phi'$,
and attach an ancilla which will be traced out to realize $\Phi(\sigma)$, $\forall \sigma\in \C C$.
The Kraus operators relate to $U$ by $K_i=\bra i| U |0\ket$ for $\{|i\ket\}$ (including $K'$)
as an orthonormal basis of the ancilla.

As an application, in the setting of QEC and computation with them,
the quantum operation after error-correction shall be a sCPTP channel on a code space
and it is the trivial channel: the identity,
and logical gates are sCPTP and often are from
unitary operators $U$ such that $[U,P]=0$.

\subsection{CP but approximate-TP maps}
\label{subsec:aCPTP}

Here we study CP maps that are not TP but approximately TP.
We denote this type of maps as aCPTP maps.
For a CP map $\Phi: \C D(\C H_1) \ra \C D(\C H_2)$ with Kraus operators $\{K_i\}$,
$\Phi^\dagger(\I)=\sum_i K_i^\dagger K_i:=K$.
Let $K=\I+\Delta$ with $K\geq 0$, $\Delta^\dagger=\Delta$.
To measure the `non-TPness' of $\Phi$,
we define a CP map $\C K$ acting on $\C D(\C H_1)$ with a single Kraus operator $\sqrt{K}$.
Then the non-TPness of $\Phi$ is that of $\C K$,
which can be measured by the distance $D_\diamond(\C K^\dagger \C K,\I)$,
the trace distance $D_t(\omega_{\C K^\dagger \C K}, \omega)$,
and the entanglement fidelity $F_E(\omega_{\C K^\dagger \C K}, \omega)$.
As they are equivalent, we will use the distance $D_\diamond$.
\begin{definition}
  [aCPTP] A CP map is $\epsilon$-TP if
$D_\diamond(\C K^\dagger \C K, \I)\leq \epsilon$.
\end{definition}
Note here since $\C K^\dagger \C K$ is not TP, both $D_t$ and $F_E$ are not upper bounded by 1.
The normalized version can be employed but not necessary.
Ignoring terms with $\Delta^2$ and higher orders,
the trace distance takes the form
\be D_t(\omega_{\C K^\dagger \C K}, \omega)=\frac{1}{2\sqrt{d_1}} \sqrt{\T tr \Delta^2}.\ee
The notion of aCPTP will be applied to the setting of quasi-exact QEC.

\section{Approximate encoding}
\label{sec:aencoding}

\subsection{AQEC with approximate encoding}
\label{sec:aqec_ae}

Here we discuss in details the extension of AQEC when the encoding is an approximate isometry.
Recall that for exact encoding, an isometry preserves orthogonality.
An encoding operation $V:\C H_L \ra \C C \subset \C H$ can be explicit or implicit:
it can be realized by a unitary circuit explicitly which maps logical states $|i\ket\in \C H_L$ to encoded states
$|\psi_i\ket \in \C C$,
or the encoded states $|\psi_i\ket \in \C C$ are prepared in a certain way
and the encoding is done implicitly, namely,
the space $\C H_L$ does not need to be carried by a physical system and is only `logical.'
Correspondingly, the explicit decoding is realized by $V^\dagger$, revealing the logical information in $\C H_L$
while the implicit decoding is realized by $P$ (when there is no error).
For both cases, an encoding isometry $V$ is a quantum channel.

Consider an approximate encoding isometry
$V: |i\ket \ra |\psi_i\ket$ and $\{|\psi_i\ket\}$ are only approximately orthogonal
while linearly independent.
The approximate projector is $P:=VV^\dagger$ and
$P^2=P+K$ for $K:=\C V(\Delta)$ and $\Delta:=V^\dagger V-\I$.
In this work, we assume the 2nd order of the overlaps $|\bra \psi_i| \psi_j\ket|$ can be ignored
in order to keep the encoding error small.
The operator $\Delta$ is specified by $\Delta_{ii}=0$, $\Delta_{ij}=\bra \psi_i| \psi_j\ket$.

The encoding $\C V$ is CP but not TP.
With the notion of aCPTP map, we say $\C V$ is $\epsilon_e$-isometric
if $d(\C V^\dagger \C V, \I)\leq \epsilon_e$.
Using trace distance of Choi states, we find
\be D_t(\omega_{\C V^\dagger \C V}, \omega)=\frac{1}{2\sqrt{d_L}}\sqrt{\sum_{i\neq j}|\bra \psi_i| \psi_j\ket|^2 }, \ee
which is a sort of average of these overlaps.

Given a logical space $\C H_L$, by simply encoding $V$ and decoding $V^\dagger$,
a small logical error will be induced if the encoding is not exact.
This is an `inherent' inaccuracy of the code itself.
We say a channel $\C N$ can be exactly detected on a code with an approximate encoding $V$ if
\be \C P \C N (\sigma)= p\; \C P (\sigma), \; p\in[0,1]. \ee
In terms of Kraus operators, the QED condition is
\be PE_iP=e_iP^2, \ee
for $\sum_i |e_i|^2=p$.
We see that this only requires the change from $P$ to $P^2$ compared with the exact encoding case.

A channel $\C N$ on a code with $\epsilon_e$-isometric encoding $V$
is $(\epsilon+\epsilon_e)$-detectable if $d(\C V^\dagger \C P \C N \C V, \C V^\dagger \C V)\leq \epsilon$.
The additivity of inaccuracy is from the triangle inequality
\be d(\C V^\dagger \C P \C N \C V, \I)\leq
d(\C V^\dagger \C P \C N \C V, \C V^\dagger \C V)+ d(\C V^\dagger \C V, \I)
\leq \epsilon+\epsilon_e. \ee

A channel $\C N$ on a code with $\epsilon_e$-isometric encoding $V$
is $(\epsilon+\epsilon_e)$-correctable if
there exists a recovery channel $\C R$ such that
$d(\C V^\dagger \C R \C N \C V,\I)\leq \epsilon+\epsilon_e$.
It is not hard to see the AQEC condition is generalized to be
\be PE_i^\dagger E_jP=a_{ij}P^2+PB_{ij}P, \ee
for $\rho_*=[a_{ij}]$, and $d(\widehat{\C N},\widehat{\C N}_0)\leq \epsilon+\epsilon_e$.
Note now the stronger condition
$d(\C V^\dagger \C R \C N \C V,\C V^\dagger \C V)\leq \epsilon$
is sufficient but not necessary.
With the above, we have shown that the approximate encoding can be easily included in the AQEC formalism.

\subsection{Transversality and quasi encoding}
\label{sec:trans_ae}

Here we discuss transversality for codes with quasi encoding.
It is known that for exact codes transversality is incompatible with universality~\cite{EK09,ZCC11,CCC+08}.
The reason is that detection of arbitrary local errors makes infinitesimal transversal operations,
which is required by universality, logically trivial.
By getting rid of infinitesimal logical gates,
it is recently proposed that
transversality is compatible with quasi universality~\cite{WZO+20}.
It is not hard to extend the argument to the setting with quasi encoding.

We first recall the `quasi-go' result.
A transversal logical gate (TLG) takes the form
\be U=\bigotimes_j U_j \label{eq:trang}\ee
for $j$ as the index of subsystems which allow independent error detection.
Given a code space $P$, it shows that the set of TLG
is a Lie group $\C G$.
For exact codes,
the connected component of identity $\C G^0$ in $\C G$ containing elements of the form
$C=\prod_k e^{i\xi_k D_k}$, $\xi_k\in \mathbb{R}$
is logically trivial, i.e., the identity gate, since each `generator' $D_k$ is logically trivial.
For quasi codes, however,
$D_k$ is not logically trivial due to the violation of the exact QEC condition, i.e.,
$D_k$ can tell the difference between codewords,
hence $\C G^0$ is not logically trivial.
Instead, $\C G^0$ is partitioned into a finite set of gate cells of various sizes
depending on the quasi code.

For the case of quasi encoding, $P^2\neq P$, there will be further perturbations from the encoding itself
in addition to perturbations by the quasi error correction (or detection).
We have to consider quasi gates which becomes exact when the encoding becomes exact, i.e., $\epsilon_e\ra 0$.
A quasi gate $U$ is defined such that
\be UP:= PUP+ L_U P,\; PU:=PUP+ P R_U, \ee
and both $L_U P$ and $P R_U$ are small and tend to zero when the encoding becomes exact.
From $P^2=P+K$, $K=V\Delta V^\dagger$,
we see that $L_{\mathds{1}} P=P R_{\mathds{1}}=-K.$
For product of two gates $U$ and $V$,
the difference between $UVP$ and $PUVP$ will increase.
Different from the exact case,
the set of quasi gates, denoted by $\C G_{\epsilon_e}$,
is an open subset of $\C G$ instead of a Lie group.
When $\epsilon_e\ra 0$, $\C G_{\epsilon_e} \ra \C G$.
We cannot employ the argument for exact encodings directly,
but we can still employ it in the following way.
A gate of the form $C=e^{i\xi D}$ can be a quasi gate, quasi identity gate, or not be a gate,
and only the first case is nontrivial.
Suppose there exists a quasi gate of the form $e^{i\xi D}$
for a certain value of $\xi$ and $D=\sum_j \alpha_j H_j$.
For other values of $\xi$, $e^{i\xi D}$ may not be a quasi gate.
Let $e^{i\xi D}P=Pe^{i\xi D}P+L_\xi P$, $\forall \xi\in \mathbb{R}$,
and $L_\xi$ may not be a smooth function of $\xi$ and may not be bounded.
We find \be DP=PDP+ A, \; A:= \lim_{\xi\ra 0} \frac{L_\xi P+K}{i\xi}.\ee
If $L_\xi$ is differentiable for all $\xi$,
then $A=-i\partial_\xi L_\xi P$, which is not a zero operator in general.
With quasi error correction, we find the difference
\be \label{eq:diff} DP- PDP= \varepsilon P^2 + PBP + A,\ee
for a certain parameter $\varepsilon$, operators $A$ and $B$.
As a result, the operator $e^{i\xi D}P$ is a nontrivial quasi gate
depending on $A$ and $B$.
This means that gates of the form $e^{i\xi D}$ are logically different.

\bibliography{ext}{}
\bibliographystyle{unsrt}

\end{document}